\documentclass[aps,prd,superscriptaddress,nofootinbib,amsmath,amsfonts,preprintnumbers,groupedaddress,showpacs,10pt,english]{revtex4-1}
\usepackage{amsmath}
\usepackage{amssymb}
\usepackage{babel}
\usepackage{wrapfig}
\usepackage{cancel}

\newcommand{\e}{\mathrm{e}}
\usepackage{relsize,exscale}
\makeatletter

\usepackage{array,multirow,graphicx}
\usepackage{dcolumn}
\usepackage{newlfont}
\usepackage{bm}
\usepackage[colorlinks,citecolor=blue,urlcolor=blue,linkcolor=blue]{hyperref}
\usepackage[figtopcap]{subfigure}
\usepackage{color}

\begin{document}

\title{Mimetic Euler-Heisenberg theory, charged solutions and multi-horizon black holes}

\author{G.~G.~L.~Nashed}
\email{nashed@bue.edu.eg}
\affiliation {Centre for Theoretical Physics, The British University, P.O. Box
43, El Sherouk City, Cairo 11837, Egypt}
\author{Shin'ichi~Nojiri}
\email{nojiri@gravity.phys.nagoya-u.ac.jp}
\affiliation{Department of Physics, Nagoya University, Nagoya 464-8602,
Japan \\
\& \\
Kobayashi-Maskawa Institute for the Origin of Particles and the Universe,
Nagoya University, Nagoya 464-8602, Japan }


\date{}

\begin{abstract}
We construct several new classes of black hole (BH) solutions in the context of the mimetic Euler-Heisenberg theory.
We separately derive three differently charged BH solutions and their relevant mimetic forms.
We show that the asymptotic form of all BH solutions behaves like a flat spacetime. { These BHs, either with/without
cosmological constant, have the non constant Ricci scalar, due to the contribution of the Euler-Heisenberg parameter,
which means that they are not solution to standard or mimetic $f(R)$ gravitational theory without the Euler-Heisenberg
non-linear electrodynamics and at the same time they are not equivalent to the solutions of the Einstein gravity
with a massless scalar field.}
Moreover, we display that the effect of the Euler-Heisenberg theory makes the singularity of BH solutions stronger compared with that of BH solutions in general relativity.
Furthermore, we show that the null and strong energy conditions of those BH solutions are violated, which is a general trend of mimetic gravitational theory.
The thermodynamics of the BH solutions are satisfactory although there appears a negative Hawking temperature under some conditions.
Additionally, these BHs obey the first law of thermodynamics.
We also study the stability, using the geodesic deviation, and derive the stability condition analytically and graphically.
Finally, for the first time and under some conditions, we derived multi-horizon BH solutions in the context of the mimetic Euler-Heisenberg theory and study their related physics.

\end{abstract}


\maketitle
\section{Introduction}\label{S1}

Einstein's general relativity (GR) constructed in (1915) \cite{Einstein:1914:PVR,Einstein:1914:FGA,Einstein:1915:ARG,Einstein:1915:EPM,Einstein:1916:GARa,2004sgig.book.....C}
is a successful and predictive theory that is considered as a cornerstone of modern physics together with quantum field theory.
The successes of GR have been confirmed multiple times, starting from gravitational lensing to the precession of Mercury's orbit \cite{Einstein:1915:EPM, Will:2005va,1920RSPTA.220..291D}.
Recently, in (2015) another GR success. i.e., the existence of gravitational waves, was confirmed by the detection of GW150914 and GW151226
by LIGO \cite{Abbott:2016blz,DeLaurentis:2016jfs,Corda:2009re}.
Despite these successes, in (1919), projects began to formulate extended GR theories such as Weyl's scale theory \cite{Weyl:1919fi} and Eddington's theory of connections \cite{dyson_1925}.
These attempts to amend GR were from the viewpoint of curiosity without any theoretical/experimental motivation.

Soon after, a theoretical framework to modify the gravitational Lagrangian of GR was created because of the non-renormalizability and thus non-quantizability of GR.
Since then, many GR modifications of GR have been proposed, e.g., $f(R)$ \cite{DeFelice:2010aj,Capozziello:2011et,Nojiri:2010wj},
$f(T)$ \cite{Capozziello:2011et,Awad:2017tyz,Nashed:2018cth} and $f(R,T)$ \cite{Harko:2011kv}.
In this study, we are interested in another modified theory, i.e., the mimetic theory.
The mimetic theory is considered an amended gravitational theory in which the conformal symmetry is viewed as an internal degree of
freedom \cite{Chaichian:2014qba,Chamseddine:2014vna,Golovnev:2013jxa,Momeni:2014qta,Deruelle:2014zza,Chaichian:2014qba}.
The expression of mimetic gravity was first introduced in \cite{Chaichian:2014qba}, and many studies on mimetic gravity
under different topics have since been carried out \cite{Nojiri:2017ncd,Sebastiani:2016ras}.
An extension of mimetic gravitational theory to $f(R)$ theory has been
formulated \cite{Nojiri:2014zqa,Odintsov:2015wwp,Odintsov:2015ocy,Odintsov:2015cwa,Leon:2014yua,Myrzakulov:2016hrx,Momeni:2015gka,Astashenok:2015haa,Myrzakulov:2015qaa}
and many cosmological applications of mimetic gravitational theory have been
conducted \cite{Cognola:2016gjy,Arroja:2015wpa,Ijjas:2016pad,Saadi:2014jfa,Matsumoto:2015wja,Myrzakulov:2015nqa,Rabochaya:2015haa}.
Moreover, several astrophysical black hole (BH) solutions in mimetic theory are discussed in
Refs.~\cite{Myrzakulov:2015kda,Oikonomou:2015lgy,Myrzakulov:2015sea,Oikonomou:2016fxb,Astashenok:2015qzw}.

The structure of mimetic theory can be explained as a specific form of a general conformal transformation, where the new and old metrics degenerate.
Using non-singular conformal transformation increases the number of degrees of freedom, and, therefore, the longitudinal mode of gravity becomes
dynamical \cite{Deruelle:2014zza,Domenech:2015tca,Firouzjahi:2018xob,Shen:2019nyp,Gorji:2019rlm}.
Usually, conformal transformation links $g_{\alpha \beta}$, which is the physical metric, to $\bar{g}_{\alpha \beta}$, (which is the auxiliary metric)
and $\psi$ (which is the scalar) by
\begin{equation}
\label{trans1}
g_{\alpha\beta}=\mp \left(\bar{g}^{\mu \nu} \partial_\mu \psi \partial_\nu \psi \right) \bar{g}_{\alpha\beta}\,.
\end{equation}
Transformation (\ref{trans1}) yields the following condition \cite{Gorji:2019rlm}
\begin{equation}
\label{trans2}
g^{\alpha \beta}\partial_\alpha \psi \partial_\beta \psi= \mp 1\,.
\end{equation}
Thus, $\partial_\alpha \psi$ is timelike and spacelike (we choose the signature of $g_{\mu\nu}$ as
$\left( g_{\mu\nu} \right) = \mathrm{diag} \left( -, +, +, + \right)$)
when we consider the negative and positive signs in (\ref{trans1}) or (\ref{trans2}), respectively.
The most well-known mimetic gravitational theory is that with the negative sign; we can consider the positive sign an extension of the theory.
An interesting feature of transformation (\ref{trans1}) is that it is non-invertible,
which means that we cannot write $\bar{g}_{\alpha \beta}$ in terms of $g_{\alpha \beta}$ \cite{Deruelle:2014zza}.
The new degree of freedom related to the transformation (\ref{trans1}) constitutes the longitudinal mode of gravity.
When we begin with the Einstein-Hilbert action that involves the physical metric $g_{\alpha \beta}$ and we apply the transformation
of Eq.~(\ref{trans1}), we obtain $\bar{g}_{\alpha \beta}$, which is the auxiliary metric, and $\psi$, which is a dynamical scalar field \cite{Chaichian:2014qba}.

The BH solutions of gravitational theory are considered as the most interesting astrophysical objects that exist in physics.
These amazing configurations supply us with a strong indication for discovering several branches of physics, e.g., thermodynamics, information theory,
paramagnetism-ferromagnetism phase transition, holographic hypothesis, superconducting phase transition, condensed matter physics,
quantum gravity, superfluids, and spectroscopy \cite{Sheykhi:2020dkm}.
Recently, BH solutions have gained much attention after several efforts to understand the confusion of the information
paradox \cite{Hawking:2016msc} and the shadow of supermassive BHs as the first results of the M87 Event Horizon Telescope \cite{Akiyama:2019cqa}.
The existence of soft hairs on the BH horizon and the complete information about their quantum state, which is stored on a holographic plate
at the horizon's future boundary, have been investigated \cite{Hawking:2016msc}.
Using these facts, one can understand the BH entropy and the microscopic structures near the horizon \cite{Afshar:2016wfy,Haco:2018ske,ElHanafy:2015egm,Haco:2019ggi,Nashed:2001im,Grumiller:2019fmp}.

The Lagrangian of non-linear electrodynamics has been constructed by Heisenberg and Euler (EH) by using the Dirac electron-positron theory \cite{Heisenberg:1935qt}.
Schwinger has reconstructed this nonperturbative one-loop Lagrangian in the frame of quantum electrodynamics (QED) \cite{schwinger1951gauge}
and has shown that this Lagrangian characterizes the phenomenon of vacuum polarization.
The imaginary part depicts the probability of the vacuum decay through the electron-positron pair production.
If electric fields are stronger than the critical value $E_c = m^2c^3/e\hbar$, the energy of the vacuum can be lowered by spontaneously creating electron-positron
pairs \cite{schwinger1951gauge,Ruffini:2013hia,
Heisenberg:1935qt}.
For a long time, experimentalists and theorists were interested in the topics of electron-positron pair production from the QED vacuum and the vacuum polarization
by an external electromagnetic field \cite{Ruffini:2009hg}.
As a cornerstone theory, QED provides a beautiful formulation of the electromagnetic interaction; additionally, it is verified experimentally.
Thence, it is urgent to study the effects of QED in black hole physics.
As a consequence of one-loop nonperturbative QED, the EH Lagrangian attracts to draw more attention to the topic of generalized BH solutions.
Many interesting BH studies in the frame of the EH non-linear electrodynamics are presented
in \cite{Bronnikov:2000vy,bronnikov1979scalar,Awad:2017sau,Yajima:2000kw,Stefanov:2007bn,Allahyari:2019jqz,Nashed:2006yw,Corda:2009xd,Ruffini:2013hia,Hendi:2014xia,Breton:2016mqh,Maceda:2018zim,Olvera:2019unw}.
Among these studies, in \cite{bronnikov1979scalar}, it has been shown that for static spherical symmetry spacetime regular BHs
are not viable for configurations with non-vanishing electric charge.
Afterward, it has been shown that this result is not trivial for dyonic configurations, where both non-zero electric and magnetic charges exist.
Moreover, it was shown that such a problem exists if we use a configuration with a pure magnetic charge.
The present aims study to derive BH solutions in the frame of the mimetic Euler-Heisenberg (MEH) theory and try to understand their physics by studying their thermodynamic properties.

The structure of this research is as follows: In Section~\ref{S2}, we give the cornerstone of the MEH theory and derive its field equations.
In Section~\ref{S3}, we apply the field equations of the MEH theory to a line element that has two unknown functions.
We study the resulting nonlinear differential equations and classify them into three cases; in each case, we derive its solution.
We show that the line-element of these solutions behaves asymptotically as flat spacetime.
We also calculate the invariants of these BHs and show that they have strong singularity compared with Einstein's GR BHs.
In Section~\ref{S3}, we study the energy conditions of each case and demonstrate that the strong and null energy conditions are violated.
In Section~\ref{S4}, we study the thermodynamics and stability of these BHs by calculating the horizons, Hawking temperature, entropy, and heat capacity.
In Subsection \ref{fir}, we show that the BHs of this study respect the first law of thermodynamics.
Finally, in Section~\ref{S6} we discuss the possibility of multi-horizon BHs and study their physical properties.
In the final Section~\ref{S66}, we summarize our findings and discuss our results.
\section{Brief summary of the MEH theory}\label{S2}

The term ``mimetic dark matter" was introduced in the literature by Mukhanov and Chamseddine \cite{Chamseddine:2014vna}, although mimetic theories had already been
constructed \cite{Lim:2010yk,Gao:2010gj,Capozziello:2010uv,Sebastiani:2016ras}. Now, we will study the MEH theory, whose action in four dimensions has the following form:
\begin{equation}
\label{act}
\mathcal{S}:=\frac{1}{2\chi} \int d^4x \sqrt{-g \left( \bar{g}_{\mu \nu},\psi \right)}R \left( \bar{g}_{\mu \nu},\psi \right)
 -\int d^4x \sqrt{-g \left( \bar{g}_{\mu \nu},\psi \right)} \mathcal{L} \left( \mathcal {F,\,G} \right)\,,
\end{equation}
where $\chi$ is the gravitational constant, $\chi\, =8\pi\,$, \, $g \left( \,\bar{g}_{\mu \nu},\psi\, \right)$ is the determinant of the metric tensor, $\psi$ is the scalar field, $R$ is
the Ricci scalar, and $\mathcal{L} \left( \mathcal{F,\,G} \right)$ is the nonlinear electromagnetic Lagrangian that
depends on invariants, $\mathcal{F}=\frac{1}{4}\mathcal{F}_{\alpha \beta}\mathcal{F}^{\alpha \beta}$
and $\mathcal{G}=\frac{1}{4}\mathcal{F}_{\alpha \beta}{^\star \mathcal{F}}^{\alpha \beta}$, where
$\mathcal{F}_{\alpha \beta}$ denotes the electromagnetic field strength tensor
and ${^\star \mathcal{F}}^{\alpha \beta}:= \frac{\sqrt{-g \left( \bar{g}_{\mu \nu},\psi \right)}}{2}\epsilon^{\alpha\beta\mu\nu} \mathcal{F}_{\mu\nu}$ is
the dual of $\mathcal{F}_{\alpha \beta}$ and $\epsilon_{\alpha \beta \gamma \delta}$ is a completely skew-symmetric tensor
that satisfies { $\epsilon^{\alpha \beta \gamma \delta} \epsilon_{\alpha \beta \gamma \delta}=-4!$}.
The Lagrangian of the Euler-Heisenberg nonlinear electromagnetic field is given by \cite{2004sgig.book.....C} 
\begin{equation}
\label{act1}
\mathcal{L}(\mathcal{F},\mathcal{G}):=-\mathcal{F}+\frac{\mu}{2}\mathcal{F}^2 +\frac{7\mu}{8}\mathcal{G}^2 \,,
\end{equation}
where $\mu=\frac{8\alpha^2}{45M^2}$ is the Euler-Heisenberg parameter
that regulates the intensity of the nonlinear electromagnetic contribution, $\alpha$ is
the fine structure constant, and $M$ is the mass of the electron,\footnote{We will take $c=\hbar=1$ in this study.} so that the Euler-Heisenberg parameter is of order
$\frac{\alpha}{{E^2}_c}$, where $E_c$ is the electric field strength.
 In the original Euler-Heisenberg theory, $\mu = \frac{8\alpha^2}{45M^2}$ is positive but
as a generalization in this paper, we also consider the case that $\mu$ is negative.

Equation~(\ref{act1}) shows that we can recover the Maxwell electrodynamics when $\mu=0$, in which $\mathcal{L}(\,\mathcal{F}\,):=-\mathcal{F}$.
In the frame of nonlinear electromagnetism, there are two possible frames,
one of which is the classical frame wherein we can use the $\mathcal{F}$ in terms
of electromagnetic field tensor $\mathcal{F}_{\alpha \beta}$.
The second frame is the $\mathcal{K}$ frame, with tensor $\mathcal{K}^{\alpha \beta}$ as the main field, which is defined by
\begin{equation}
\label{act2}
\mathcal{K}^{\alpha \beta}:=-\left(\mathcal{L}_\mathcal{F}\mathcal{F}^{\alpha \beta}+{^\star \mathcal{F}}^{\alpha \beta}\mathcal{L}_\mathcal{G}\right)\,,
\end{equation}
where $\mathcal{L}_\mathcal{F}$ is the derivative of $\mathcal{L}$ w.r.t. $\mathcal{F}$ and $\mathcal{L}_\mathcal{G}$ is the derivative of $\mathcal{L}$ w.r.t $\mathcal{G}$.
In the MEH, $\mathcal{K}_{\alpha \beta}$ takes the following form,
\begin{equation}
\label{act11}
\mathcal{K}^{\alpha \beta}:=(1-\mu \mathcal{F})\mathcal{F}^{\alpha \beta}-\frac{7\mu}{4}\,\,{^\star \mathcal{F}}^{\alpha \beta}\mathcal{G} \,.
\end{equation}
The aforementioned tensor corresponds to electric induction $D$ and
magnetic field $H$, and Eq.~(\ref{act11}) shows the constitutive relations
between $D$, $H$, magnetic intensity $B$, and electric
field $E$ in the Euler-Heisenberg nonlinear electromagnetism.

The two invariants, $\mathcal{K}$ and $\mathcal{O}$, associated with
the $\mathcal{K}$ frame are given by
\begin{equation}
\label{act111}
\mathcal{K}:=-\frac{1}{4}\mathcal{K}_{\alpha \beta}\mathcal{F}^{\alpha \beta} \,, \qquad
\mathcal{O}:=-\frac{1}{4}\mathcal{K}_{\alpha \beta}{^\star \mathcal{K}}^{\alpha \beta} \,,
\end{equation}
where ${^\star \mathcal{K}}_{\alpha \beta}=\frac{1}{2\sqrt{-g \left( \bar{g}_{\mu \nu},\psi \right)}}\epsilon_{\alpha \beta \gamma \delta}\mathcal{K}^{\gamma \delta}$.

Using the aforementioned information, we can obtain the field equations of MEH in the form
\begin{equation} \label{fe3}
I_\alpha{^\beta}\equiv G_\alpha{^\beta}-{\mathcal{T}^\mathrm{EH}}_\alpha{^\beta}-\widetilde{\mathcal{T}}_\alpha{}^ \beta= 0, \qquad \qquad \nabla_\alpha\,\mathcal{K}^{\alpha \beta}=0\,,
\end{equation}
where $ G_{\alpha \beta}$ is the Einstein tensor, $\widetilde{\mathcal{T}}_{\alpha}{^\beta}$ is the energy-momentum tensor of the mimetic field
and ${\mathcal{T}^\mathrm{EH}}{_\alpha}{^\beta}$ is the energy-momentum tensor of MEH in the $\mathcal{K}$ frame that takes the form:
\begin{equation}
\label{maxf1}
{\mathcal{T}^\mathrm{EH}}_{\alpha}{^\beta}=\frac{1}{4\pi}\left( \left[ 1-\mu \mathcal{K} \right] \mathcal{K}^\mu{}_\alpha\,\mathcal{K}_{\beta \mu}
+\delta_\alpha{}^\beta \left[ \mathcal{K}-\frac{3}{2}\,\mu\mathcal{K}^2-\frac{7\mu}{8}\,\mathcal{O}^2\right]\right)\, .
\end{equation}
Notably, the auxiliary metric in the field equations, appears implicitly through the physical metric given in Eq.~(\ref{trans2}) and mimetic field $\psi$.
The presence of the mimetic field in the field equations can be written as
\begin{equation}
\label{ten1}
\widetilde{\mathcal{T}}_{\alpha \beta}=-\left(G-{\mathcal{T}^\mathrm{EH}}\right)\partial_\alpha\psi\, \partial_\beta\psi,\,
\end{equation}
where $G = -R$ is the trace of the Einstein tensor.
It is worth mentioning that energy-momentum tensors, ${\mathcal{T}^\mathrm{EH}}_{\mu \nu}$ and $\widetilde{\mathcal{T}}_{\mu \nu}$ ,
are conserved, i.e., satisfy the continuity equations $\nabla^\mu {\mathcal{T}^\mathrm{EH}}_{\mu \nu}=0=\nabla^\mu \widetilde{\mathcal{T}}_{\mu \nu}$,
where $\nabla^\mu$ is the covariant derivative. Using the mimetic field constraint (\ref{trans2}) and the energy-momentum tensor (\ref{ten1}),
the corresponding continuity reads
\begin{equation}
\label{cons3}
\nabla^\kappa\left(\left[G-{\mathcal{T}^\mathrm{EH}}\right]\partial_\kappa \psi\right)= \frac{1}{\sqrt{-g}}\partial_\kappa
\left(\sqrt{-g}\left[G-{\mathcal{T}^\mathrm{EH}}\right] g^{\kappa \sigma}\partial_\sigma \psi \right)=0\, ,
\end{equation}
 which is also obtained by the variation of the action (\ref{act}) w.r.t. the mimetic scalar field $\psi$.

Alternatively, it can be seen that (\ref{cons3}) is satisfied identically, when (\ref{trans2}) is used.
It is straightforward to show that the trace of Eq.~(\ref{fe3}) has the form
\begin{equation}
\left[G-{\mathcal{T}^\mathrm{EH}}\right]\left(1+g^{\mu \nu}\partial_\mu\psi \partial_\nu \psi\right)\equiv 0\, ,
\end{equation}
which is satisfied identically because of the mimetic field constraint, namely Eq.~(\ref{trans2}).
In conclusion, we note that the conformal degree of freedom provides a dynamical quantity, i.e., ($G \neq 0$), therefore,
the mimetic theory has nontrivial solutions for the conformal mode even in the absence of matter \cite{Chamseddine:2013kea}.

\section{Charged MEH BH solutions}\label{S3}

In this section, we apply the Euler-Heisenberg field equations of the mimetic theory to a spherically symmetric spacetime and try to solve them.
For this purpose, we use the following metric:
\begin{equation}
\label{met}
ds^{2}=f(r)dt^{2}-\frac{dr^{2}}{f_1(r)}-r^2\left(d\theta^{2}+r^2\sin^2\theta d\phi^{2}\right)\,,
\end{equation}
where $f(r)$ and $f_1(r)$ are unknown functions, which we will determine by using the field equations.
Applying the MEH field equations to spacetime (\ref{met}), we obtain the following nonlinear differential equations: \\
The $(t,t)$-component of the MEH equation is
\begin{align}
0=&\frac{128f^6[1-rf_1'-f_1]-q'^2f_1r^2[64f^5+\mu (48q'^2f_1f^4-16\mu f^3f_1^2q'^4-24\mu^2f^2f_1^3q'^6-8\mu^3ff_1{}^4q'^8-\mu^4f_1{}^5q'^{10})]}{128r^2f^6}\,, \nonumber
\end{align}
the $(r,r)$-component of the MEH equation is
\begin{align}
0=&\frac{1}{128r^2f^6}\left(128f^5 \left[ f-ff_1-rf_1f' \right]-4\psi'^2f_1 \left[32f_1f^5r^2f''+q'^2r^2\mu \left(
16q'^2f_1f^4+32\mu f^3f_1^2q'^4
\right. \right. \right. \nonumber\\
& \left. \left. +24\mu^2f^2f_1^3q'^6+8\mu^3ff_1{}^4q'^8
+\mu^4f_1{}^5q'^{10} \right) +16f^4 \left( rff' \left[ 4f_1+rf_1' \right]-r^2f_1f'^2+4f^2 \left[ f_1-1+rf'_1 \right] \right) \right] \nonumber\\
& \left. -q'^2f_1r^2 \left[ 64f^5 +\mu \left(48q'^2f_1f^4-16\mu f_1{}^2f^3q'^4
 -24\mu^2f^2f_1^3q'^6-8\mu^3ff_1{}^4q'^8-\mu^4f_1{}^5q'^{10} \right) \right] \right) \,,\nonumber
\end{align}
both of the $(\theta,\theta)$ and $(\phi,\phi)$-components of the MEH equation give the following equation,
\begin{align}
\label{fe}
0=& \frac{1}{128rf^6} \left( 32f^4 \left[ rf_1f'^2-ff' \left[ 2f_1+rf_1' \right] r^2-2f^2f_1' \right] -64rf^5f_1f''+q'^2f_1r \left[ 64f^5
\right. \right. \nonumber\\
& \left. \left. + \mu \left(16q'^2f^4-80\mu f^3f_1^2q'^4-72\mu^2f^2f_1^3q'^6
 -24\mu^3ff_1{}^4q'^8-3\mu^4f_1{}^5q'^{10} \right) \right] \right) \,,
\end{align}
and the nonlinear charged field equation takes the following form
\begin{equation}
\label{chr}
0= -\frac{2rf_1fq'' \left( 2f+3\mu f_1q'^2 \right) + q' \left( q'^2\mu f_1 \left\{ 3 rff_1'+f_1 \left[ 4f-3rf' \right] \right\}
+2f \left[ rff_1'+f_1 \left( 4f-rf' \right) \right] \right)}{4rf^3}=0\,,
\end{equation}
where $q\,\equiv q(r)$
is an unknown function related to the electric field which is defined by vector potential
\begin{equation}
\label{chr1}
V = q(r)dt\,,
\end{equation}
and $q'=\frac{dq}{dr}$.
We will solve the aforementioned nonlinear differential equations, i.e., (\ref{fe}) and (\ref{chr})
for the following three different cases:

\

\noindent
\underline{Case I: $f(r)=f_1(r)$}\vspace{0.2cm}\\
When $f(r)=f_1(r)$ the analytic solution of the nonlinear differential equations, (\ref{fe}) and (\ref{chr}) takes the following forms
\begin{align}
\label{sol1}
f(r)=& 1+\frac{c_1}{r}+\frac{c_2{}^2}{8r^2}-\frac{c_2{}^4\mu}{640 r^6}\,, \nonumber\\
q(r)=& \mathlarger{\mathlarger{\mathlarger{\mathlarger{\int}}}}\frac{2\sqrt[3]{18r^2\mu^3\left(9c_2\sqrt{\mu}+\sqrt{3(32r^4+27c_2{}^2\mu)} \right)^2}-24r^2\mu}
{6r\mu\sqrt[3]{r\mu\left(108c_2 \mu+12\sqrt{3\mu \left( 32r^4+27c_2{}^2\mu \right)}\right)}}dr+c_3\nonumber\\
\approx& c_3-\frac{c_2}{2r}+\frac{\mu c_2{}^3}{80r^5}-\frac{\mu^2 c_2{}^5}{384r^9}+\frac{3\mu^3 c_2{}^7}{3328r^{13}}-\mathcal{O}\left(\frac{1}{r^{17}}\right)+\cdots\,.
\end{align}
Equation~(\ref{sol1}) shows that when nonlinear parameter $\mu=0$, we return to the well-known charged BH of GR, i.e., the Reissner--Nordstr\"om solution.
Using Eq.~(\ref{sol1}), we obtain the mimetic field in the form
\begin{eqnarray}
\label{scal1}
\psi=8\mathlarger{\mathlarger{\mathlarger{\int}}}\sqrt{\frac{10r^6}{80c_2{}^2r^4-\mu c_2{}^4+640r^5 \left( r+c_1 \right)}}dr
\approx r-\frac{1}{2}c_1\ln(r)+\frac{c_2{}^2-6c_1{}^2}{16r}+\frac{c_1 \left( 10c_1{}^2-3c_2{}^2 \right)}{64r^2} + \mathcal{O}\left(\frac{1}{r^{3}}\right)+\cdots\,.\nonumber\\
&&
\end{eqnarray}
Using Eq.~(\ref{sol1}) in (\ref{met}), we obtain the line element in the following form
\begin{eqnarray}
\label{elm1}
ds^2=\left(1-\frac{2M}{r}+\frac{q^2}{r^2}-\frac{q^4\mu}{10 r^6}\right)dt^2-\frac{dr^2}{1-\frac{2M}{r}+\frac{q^2}{r^2}-\frac{q^4\mu}{10 r^6}}
 -r^2 \left(d\theta^2 + \sin^2\theta d\phi^2 \right)\,,
\end{eqnarray}
where $c_1=-2M$ and $c_2=\sqrt{8}q$.

Using Eq.~(\ref{elm1}), we get the invariants of solution (\ref{sol1}) as:
\begin{align}
\label{inv}
\mathcal{R}_{\mu \nu \rho \sigma} \mathcal{R}^{\mu \nu \rho \sigma}=&\frac{48M^2}{r^6}-\frac{96Mq^2}{r^7}+\frac{56q^4\mu }{r^8}
+\frac{224Mq^4\mu }{5r^{11}}-\frac{304q^6\mu }{5r^{12}}+\frac{478q^8\mu^2 }{25r^{16}}\,, \nonumber\\
\mathcal{R}_{\mu \nu } \mathcal{R}^{\mu \nu }=& \frac{4q^4}{r^8}-\frac{8q^6\mu}{r^{12}}+\frac{5q^8\mu^2 }{r^{16}}\,, \qquad \mathcal{R}=-\frac{2q^4\mu}{r^8}\,.
\end{align}
Here $\mathcal{R}_{\mu \nu \rho \sigma} \mathcal{R}^{\mu \nu \rho \sigma}$,
$\mathcal{R}_{\mu \nu} \mathcal{R}^{\mu \nu}$, and $\mathcal{R}$ represent
the Kretschmann scalar, the Ricci tensor square, and the Ricci scalar, respectively and they all have
a true singularity when $r=0$. It is important to stress that the parameter $\mu$
is the origin of the differences from the aforementioned results in the Reissner-Nordstr\"om BH
of GR, whose invariants behave as $\left( \mathcal{R}_{\mu \nu \rho \sigma}
\mathcal{R}^{\mu \nu \rho \sigma}, \mathcal{R}_{\mu \nu} \mathcal{R}^{\mu \nu},
\mathcal{R} \right) = \left( \frac{48M^2}{r^6}, \frac{4q^4}{r^8}, 0 \right)$.
Equation~(\ref{inv}) shows that the leading term of variants $\left(
\mathcal{R}_{\mu \nu \rho \sigma} \mathcal{R}^{\mu \nu \rho \sigma},\mathcal{R}_{\mu \nu}
\mathcal{R}^{\mu \nu},\mathcal{R} \right)$ is the same as that of the Reissner-Nordstr\"om BH, except for Ricci scalar, whose leading term behaves as
$\mathcal{O}\left(\frac{1}{r^{8}}\right)$.

\

\noindent
\underline{Case II: $f(r)\neq f_1(r)$}\\
When $f(r)\neq f_1(r)$, it is difficult to solve differential equations, (\ref{fe}) and (\ref{chr}) explicitly.
Even in this case $f(r)\neq f_1(r)$, we can use the expression of $q(r)$ in (\ref{sol1}) and expand $q(r)$ w.r.t. $1/r$ as
in the expression of the last line in (\ref{sol1}) and we keep first four terms as follows,
\begin{eqnarray}
\label{sol2char1}
q(r)= c_3-\frac{c_2}{2r}+\frac{\mu c_2{}^3}{80r^5}-\frac{\mu^2 c_2{}^5}{384r^9} + \mathcal{O}\left( r^{\frac{1}{13}} \right)\, .
\end{eqnarray}
We should note that we cannot put $\mu=0$ in the expression of $q(r)$ in (\ref{sol1}) but by expanding the expression w.r.t. $1/r$,
we can reproduce the form of the Maxwell theory with $\mu=0$ asymptotically.

Using Eq.~(\ref{sol2char1}) in Eq.~(\ref{fe}), we get
\begin{align}
\label{sol2}
f(r)=&1+\frac{c_1}{r}+\frac{c_2{}^2}{8r^2}-\frac{c_2{}^4\mu}{640 r^6}+\frac{9\mu^2 c_2{}^4}{16 r^{8}}+\frac{9\mu^2c_1 c_2{}^4}{16 r^{9}}+\frac{\mu^2 c_2{}^6}{16 r^{10}}\,, \nonumber\\
f_1(r)=&1+\frac{c_1}{r}+\frac{c_2{}^2}{8r^2}-\frac{\mu c_2{}^4}{640 r^6}-\frac{\mu^2 c_2{}^6}{128 r^{10}}\,.
\end{align}
Equation~(\ref{sol2}) shows that when $\mu=0$, we return to the well-known charged BH of GR, i.e., the Reissner--Nordstr\"om solution.
Using Eq.~(\ref{sol2}), we obtain the mimetic field in the form
\begin{align}
\label{scal1x}
\psi=&\mathlarger{\mathlarger{\mathlarger{\int}}}\frac{80r^{5}}{\sqrt{6400r^{10}+800c_2{}^2r^8-10\mu c_2{}^4r^4-50\mu^2c_2{}^6+6400c_1r^9}}dr\nonumber\\
\approx& r-\frac{1}{2}c_1\ln(r)+\frac{c_2{}^2-6c_1{}^2}{16r}-\frac{c_1 \left( 10c_1{}^2-3c_2{}^2 \right)}{64r^2}+ \mathcal{O}\left(\frac{1}{r^{3}}\right)+\cdots\,.
\end{align}
Using Eq.~(\ref{sol2}) in (\ref{met}) we get the line element in the form
\begin{align}
\label{elm2}
ds^2=&\left(1-\frac{2M}{r}+\frac{q^2}{r^2}-\frac{q^4\mu}{10 r^6}+\frac{36q^4\mu^2}{ r^8}-\frac{72Mq^4\mu^2}{ r^9}+\frac{32q^6\mu^2}{ r^{10}}\right)dt^2
 -\frac{dr^2}{1-\frac{2M}{r}+\frac{q^2}{r^2}-\frac{q^4\mu}{10 r^6}-\frac{4q^6\mu^2}{r^{10}} }\nonumber\\
& -r^2 \left( d\theta^2 + \sin^2\theta d\phi^2 \right) \, ,
\end{align}
where $c_1=-2M$ and $c_2=\sqrt{8}q$ as in Case I.
Equation~(\ref{elm2}) shows also that when the parameter $\mu=0$, we obtain the Reissner--Nordstr\"om metric.
If we follow the same procedure as in Case I to calculate the invariants of line-element (\ref{elm2}),
we find $\left( \mathcal{R}_{\mu \nu \rho \sigma}
\mathcal{R}^{\mu \nu \rho \sigma}, \mathcal{R}_{\mu \nu} \mathcal{R}^{\mu \nu},
\mathcal{R} \right) = \left( \frac{48M^2}{r^6}, \frac{4q^4}{r^8}, 0 \right)$ as in Case I.


\

\noindent
\underline{Case III: $f_1(r)= f(r)f_2(r)$}\\

When $f_1(r)= f(r)f_2(r)$, we use the expression $q(r)$ in (\ref{sol2char1}), again.
By substituting Eq.~(\ref{sol2char1}) into (\ref{fe}), we obtain
\begin{align}
\label{sol3}
f(r) =& 1+\frac{c_1}{r}+\frac{c_2{}^2}{8r^2}-\frac{\mu c_2{}^4}{640 r^6}-\frac{9\mu^2 c_2{}^4}{112 r^8}-\frac{35\mu^3 c_2{}^6}{1408 r^{12}}\,, \nonumber\\
f_2(r) =& 1-\frac{9\mu^2c_2{}^4}{16r^8}-\frac{35\mu^3 c_2{}^6}{128r^{12}}\,.
\end{align}
Equation~(\ref{sol3}) shows that when $\mu=0$, we return to the well-known charged BH of GR, i.e., the Reissner--Nordstr\"om solution, again.
Using Eq.~(\ref{sol3}) we get the mimetic field in the form
\begin{align}
\label{scal3}
\psi=& \nonumber\\
%
\mathlarger{\int} &
\frac{128c_1\sqrt{385}r^{12} dr}{\sqrt{
\left( 77c_2^6\mu r^6-49280c_2^2r^{12}-6160c_4^2r^{10}+3960c_2^6\mu^2 r^4+1225c_2^8\mu^3-49280c_1c_2^2r^{11} \right)
\left( 35\mu^3 c_2^6-128r^{12}+72\mu^2c_2^4r^4 \right)}} \nonumber\\
\approx& r-\frac{1}{2}c_1\ln(r)-\frac{c_2{}^2-2c_1{}^2}{16r}-\frac{c_1(2c_1{}^2-c_2{}^2)}{64r^2}+ \mathcal{O}\left(\frac{1}{r^{3}}\right)+\cdots\,.
\end{align}
Equation~(\ref{sol3}) shows that when $\mu=0$, we obtain $f_2=1$ and, hence, return to
 the case of $f_1(r)= f(r)$ in Case I.

Using Eq.~(\ref{sol3}) in (\ref{met}), we find the line element in the form
\begin{align}
\label{elm3}
ds^2=&\left(1-\frac{2M}{r}+\frac{q^2}{r^2}-\frac{q^4\mu}{10 r^6}-\frac{36q^4\mu^2}{7r^8}-\frac{140q^6\mu^3}{11r^{12}}\right)dt^2\nonumber\\
& -\frac{dr^2}{\left(1-\frac{2M}{r}+\frac{q^2}{r^2}-\frac{q^4\mu}{10 r^6}-\frac{36q^4\mu^2}{7r^8}+\frac{140q^6\mu^3}{11 r^{12}}\right)
\left(1-\frac{36q^4\mu^2}{r^8}-\frac{140q^6\mu^3}{r^{12}}\right)}-r^2 \left( d\theta^2 + \sin^2\theta d\phi^2 \right)\,.
\end{align}
Equation~(\ref{elm3}) also shows also that when parameter $\mu=0$, we obtain the Reissner--Nordstr\"om metric.
If we follow the same procedure as in Case I to calculate the invariants of line-element (\ref{elm2}),
we find the same asymptotic values as Case I,
 $\left( \mathcal{R}_{\mu \nu \rho \sigma}
\mathcal{R}^{\mu \nu \rho \sigma}, \mathcal{R}_{\mu \nu} \mathcal{R}^{\mu \nu},
\mathcal{R} \right) = \left( \frac{48M^2}{r^6}, \frac{4q^4}{r^8}, 0 \right)$.
{ In the next subsection we are going to derive a charged solution with a cosmological constant.}

{
\subsection{Charged BH with cosmological constant}\label{lambda}
In this subsection, we are going to derive a spherically symmetric charged BH with cosmological constant in the frame MEH.
For this aim, we write the field equations (\ref{fe3}) with a cosmological constant that yields the form:
\begin{equation}
\label{fe4}
I_{\alpha \beta}\equiv G_{\alpha \beta}-2\Lambda g_{\alpha \beta}-{\mathcal{T}^\mathrm{EH}}_{\alpha \beta}-\widetilde{\mathcal{T}}_{\alpha \beta}= 0\, , \quad
\nabla^\alpha\,\mathcal{K}_{\alpha \beta}=0\,,
\end{equation}
where $\Lambda$ is the cosmological constant.

Applying the MEH field equations (\ref{fe4}) to spacetime (\ref{met}), we obtain the following nonlinear differential
equations\footnote{Here we use the metric (\ref{met}) with $f=f_1$ and other cases can be followed same as the case of $f=f_1$.}: \\
The $(t,t)$-component of the MEH equation is
\begin{align}
0=&\frac{128[1-rf'-f]-256\Lambda r^2-q'^2r^2[64+\mu (48q'^2-16\mu q'^4-24\mu^2q'^6-8\mu^3q'^8-\mu^4q'^{10})]}{128r^2}\,, \nonumber
\end{align}
the $(r,r)$-component of the MEH equation is
\begin{align}
0=&\frac{1}{128r^2}\left(128 \left[ 1-f-rf' \right]-256\Lambda r^2 [1+4\psi'^2f]-4\psi'^2 \left[32r^2f''+q'^2r^2\mu \left(
16q'^2+32\mu q'^4 +24\mu^2q'^6+8\mu^3q'^8 \right. \right. \right. \nonumber\\
& \left. \left. +\mu^4q'^{10} \right) +64 \left( 2rf'-1+f \right) \right] -q'^2r^2 \left[ 64 +\mu \left(48q'^2-16\mu q'^4
 -24\mu^2q'^6-8\mu^3q'^8-\mu^4q'^{10} \right) \right] \,,\nonumber
\end{align}
both of the $(\theta,\theta)$ and $(\phi,\phi)$-components of the MEH equation give the following equation,
\begin{align}
\label{fee}
0=& \frac{1}{128r} \left(64\left[2f'+rf'' \right] +256\Lambda r^2-q'^2r \left[ 64
+ \mu \left(16q'^2-80\mu\,q'^4-72\mu^2\,q'^6 -24\mu^3\,q'^8-3\mu^4q'^{10} \right) \right] \right) \,,
\end{align}
and the nonlinear charged field equation takes the following form
\begin{equation}
\label{chrr}
0= -\frac{2rq''+3rq''q'^2\mu+3q'+2\mu\,q'^3}{2r}=0\,.
\end{equation}

In the system of Eqs.~(\ref{fee}) and (\ref{chrr}), if we set $\Lambda=0$, we return to the Case I discussed aforementioned.
If we set $\Lambda\neq 0$ and in the absence of the EH parameter $\mu=0$, we already get the Reissner-Nordstr\"om solution
where the Ricci scalar has a constant value, i.e., $R=4\Lambda$.
However, when $\Lambda\neq$, $\psi \neq 0$, and $\mu\neq 0$, we get the following solution to the system of Eqs.~(\ref{fee}) and (\ref{chrr})
\begin{align}
\label{sol1}
f(r)=& 1-\frac{2\Lambda r^2}{3}-\frac{2M}{r}+\frac{q{}^2}{8r^2}-\frac{q{}^4\mu}{10 r^6}\,, \nonumber\\
q(r)=& \frac{1}{\sqrt{3}\mu} \int \!{\frac {\sqrt {\left( 4{r}^{4}+54\,\mu\,q^2+6q\sqrt {3\mu[4{r}^{4}
+27\mu\,q^2]} \right) ^{2/3}+2\sqrt [3]{2r^8}-4\sqrt [3]{2\,{r}^{8}
+27\,\mu\,q^2\,r^4+3r^4q\sqrt {3\mu[4{r}^{4}+27\mu\,q^2]}}}}{\sqrt [6]{2\,{r}^{4}+27\,\mu\,q^2+3q\sqrt{3\mu[4{r}^{4}
+27\mu\,q^2]}}{r}^{2/3}}}{dr}+c_3\,,
\nonumber\\
\approx& c_3-\frac{c_2}{2r}+\frac{\mu c_2{}^3}{80r^5}-\frac{\mu^2 c_2{}^5}{384r^9}+\frac{3\mu^3 c_2{}^7}{3328r^{13}}
 -\mathcal{O}\left(\frac{1}{r^{17}}\right)+\cdots\,,\nonumber\\
\psi=&\int \!\sqrt {{\frac {30{r}^{6}}{30\,{r}^{6}-20\,{r}^{8}{\Lambda}-60M\,{r}^{5}+30q^2\,{r}^{4}
 -3\mu\,{q}^{4}}}}{dr}\approx \sqrt {\frac{3}{2|\Lambda|}}\ln\,r
-{\frac {3}{16\,r^2}}\sqrt {\frac{6}{|\Lambda^3|}}+{\frac {1}{4\,r^3}}\sqrt {\frac{6}{|\Lambda^3|}}+\cdots\,.
\end{align}
Equation~(\ref{sol1}) shows that when nonlinear parameter $\mu=0$, we return to the well-known charged BH of GR, i.e.,
the Reissner--Nordstr\"om AdS/dS spacetime.

In GR, and when $\mu=0$, the Ricci scalar has a constant value $R=-8\Lambda$, however,
when $\mu\neq 0$, $R=-8\Lambda-\frac{2q^2\mu}{r^8}$, which means that we have a non-constant Ricci scalar.
In the mimetic gravity without EH parameter Ricci scalar has a constant value which means that the BH solution (\ref{sol1}) will be a solution
for $f(R)$ mimetic theory.
However, when the EH parameter involved in the field equation, we have a solution that yields a non-constant Ricci scalar solution, which means
that the BH (\ref{sol1}) is not a solution $f(R)$ mimetic theory.
The main reason for this is the contribution of the EH parameter.
This case needs more study in the frame of $f(R)$ mimetic theory to prove this statement.}

\subsection{Energy conditions}

\begin{figure*}
\centering
\subfigure[~WEC]{\label{fig:1a}\includegraphics[scale=0.2]{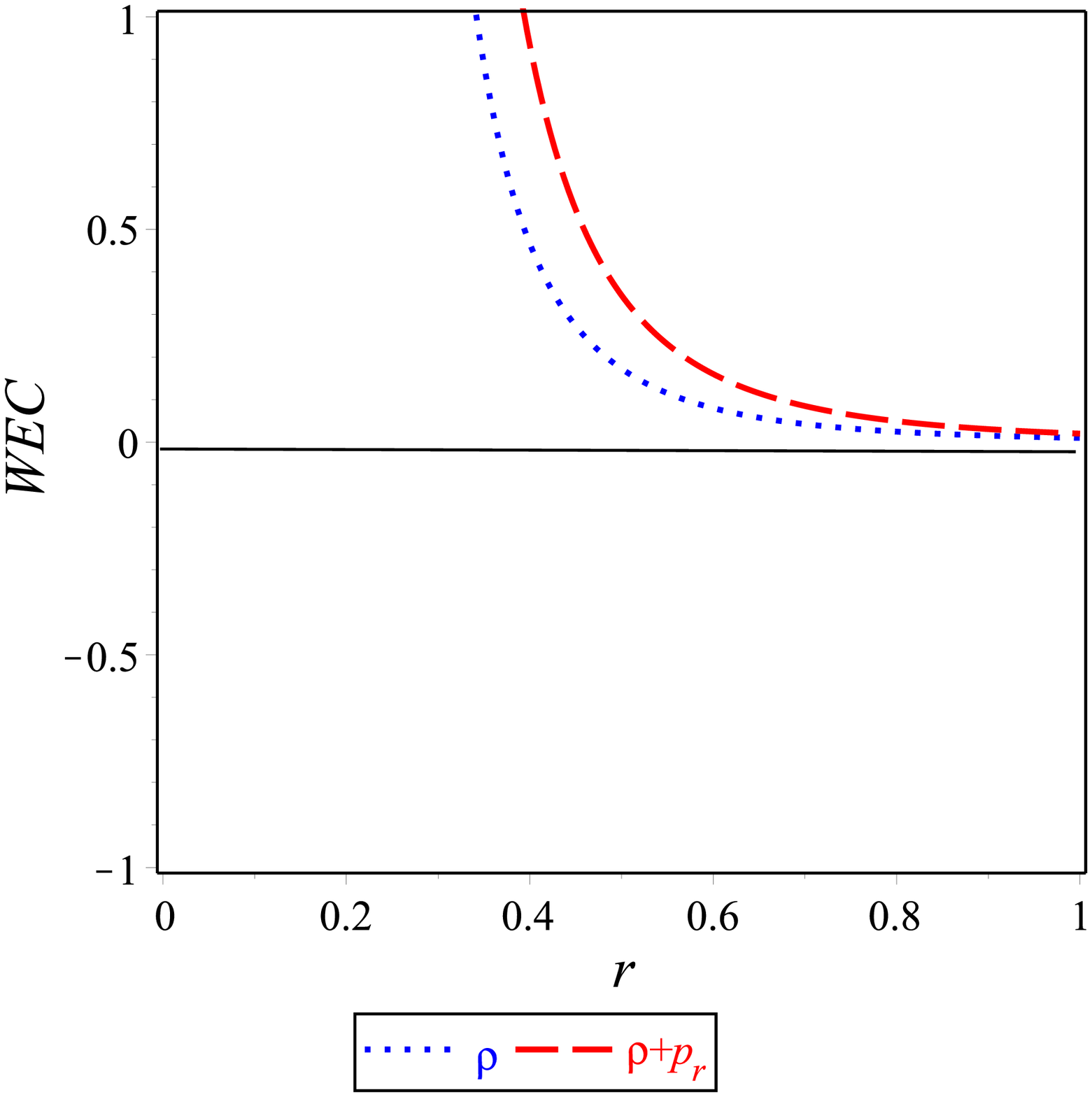}}\hspace{0.5cm}
\subfigure[~SEC]{\label{fig:1b}\includegraphics[scale=0.2]{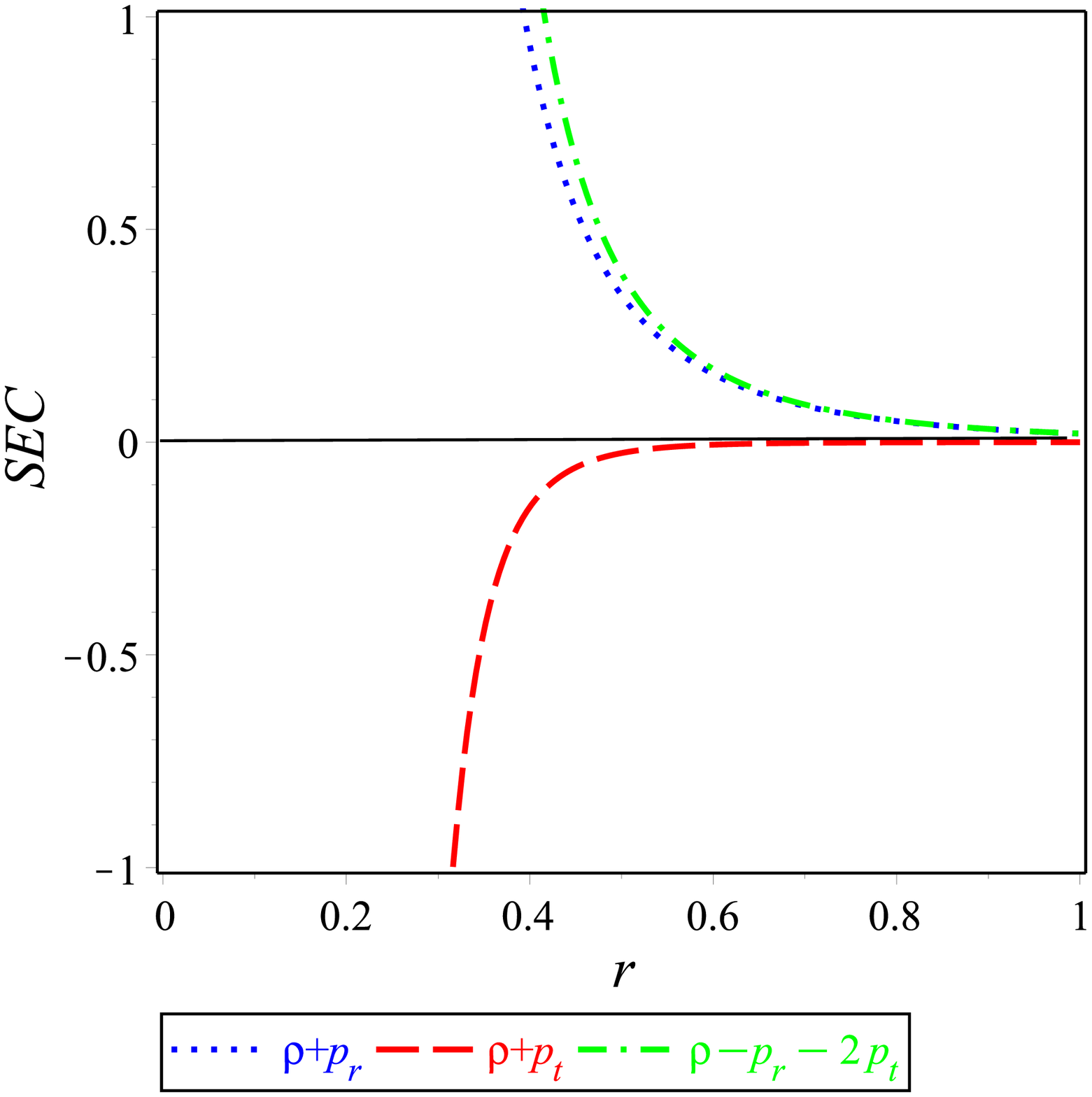}}\hspace{0.5cm}
\subfigure[~NEC]{\label{fig:1c}\includegraphics[scale=0.2]{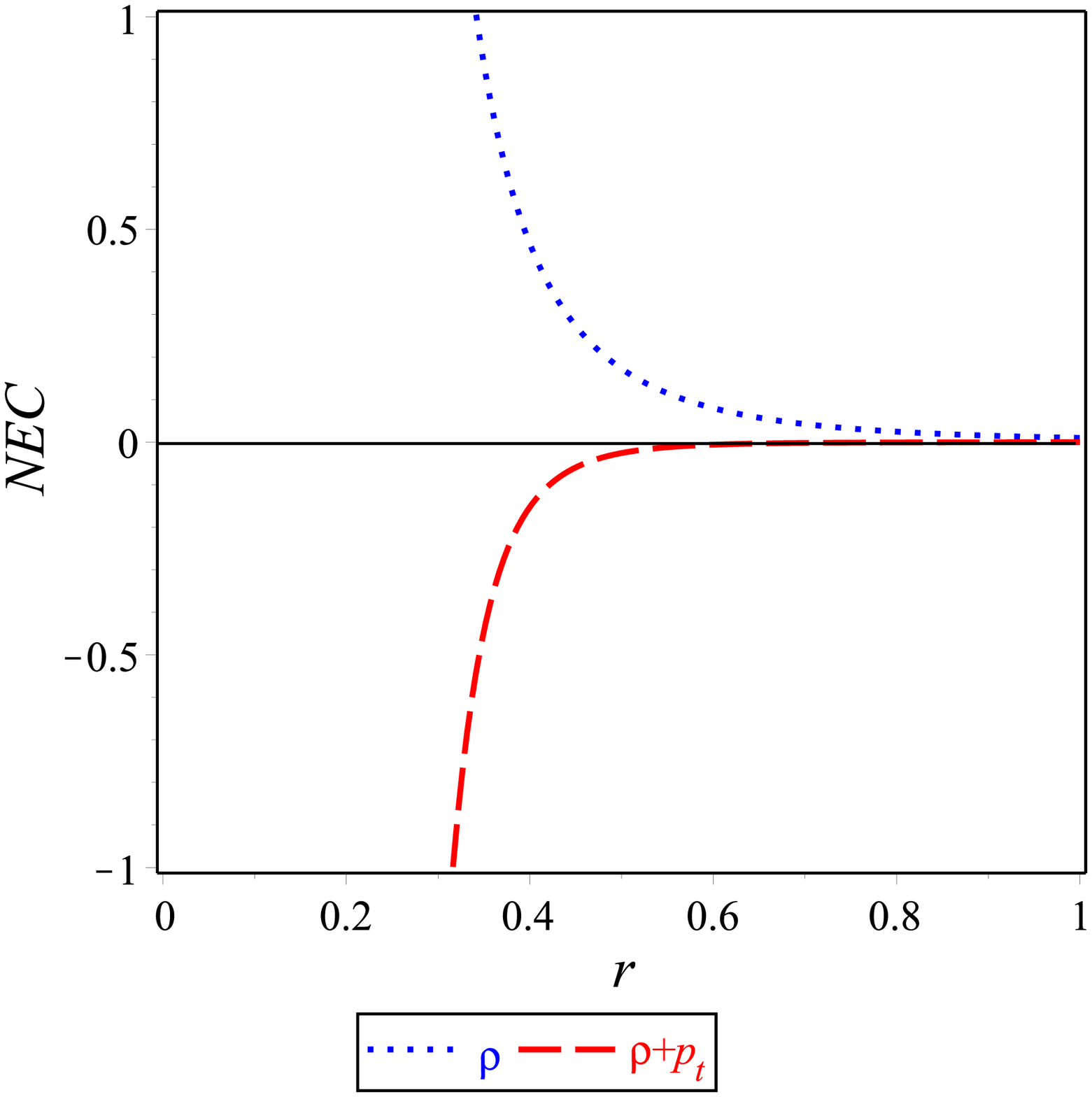}}\hspace{0.5cm}
\subfigure[~DEC]{\label{fig:1d}\includegraphics[scale=0.2]{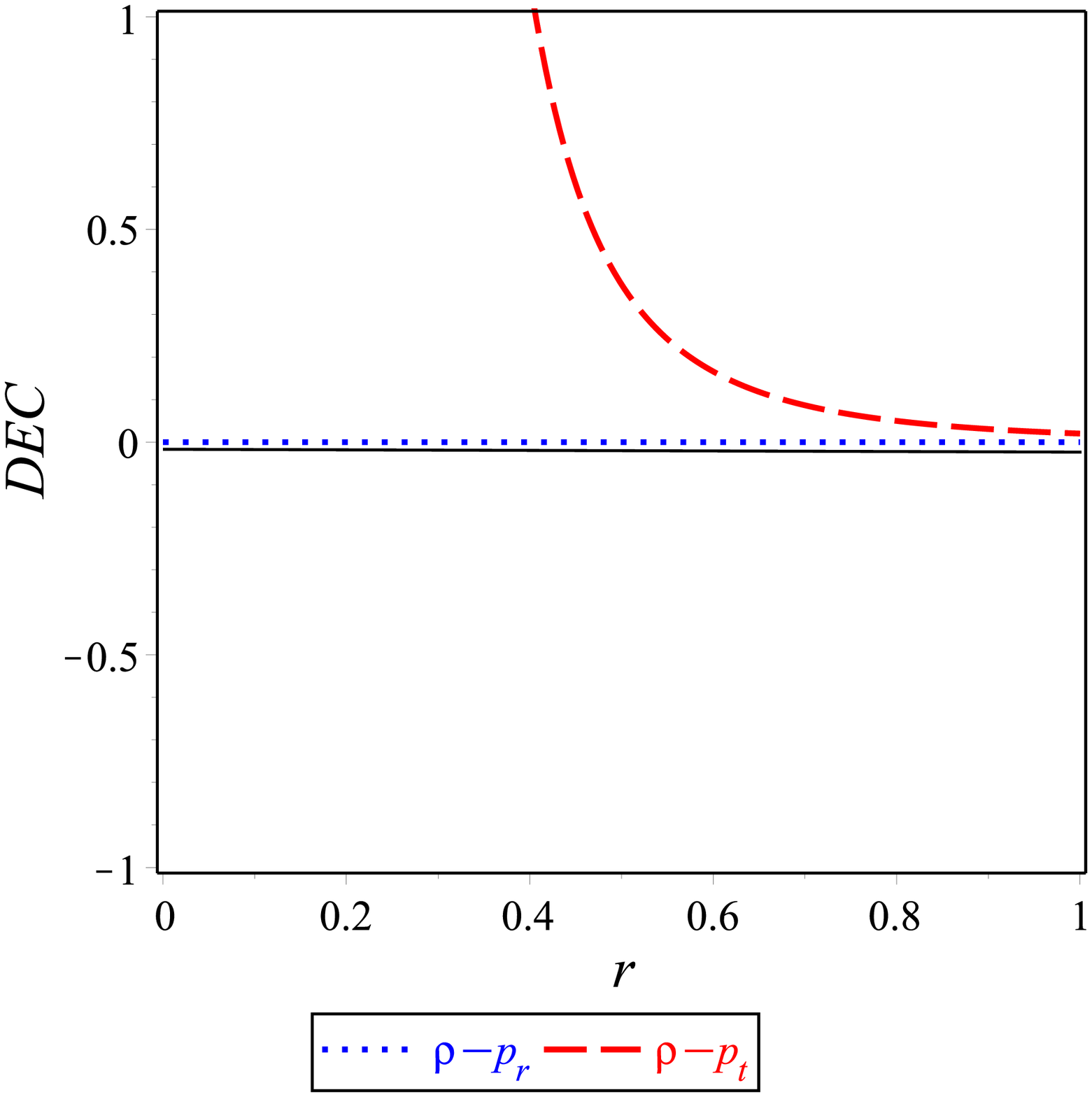}}
\caption{Schematic plots of WEC, SEC, NEC and DEC given by Eq.~(\ref{elm1}).}
\label{Fig:1}
\end{figure*}

\begin{figure*}
\centering
\subfigure[~WEC]{\label{fig:2a}\includegraphics[scale=0.2]{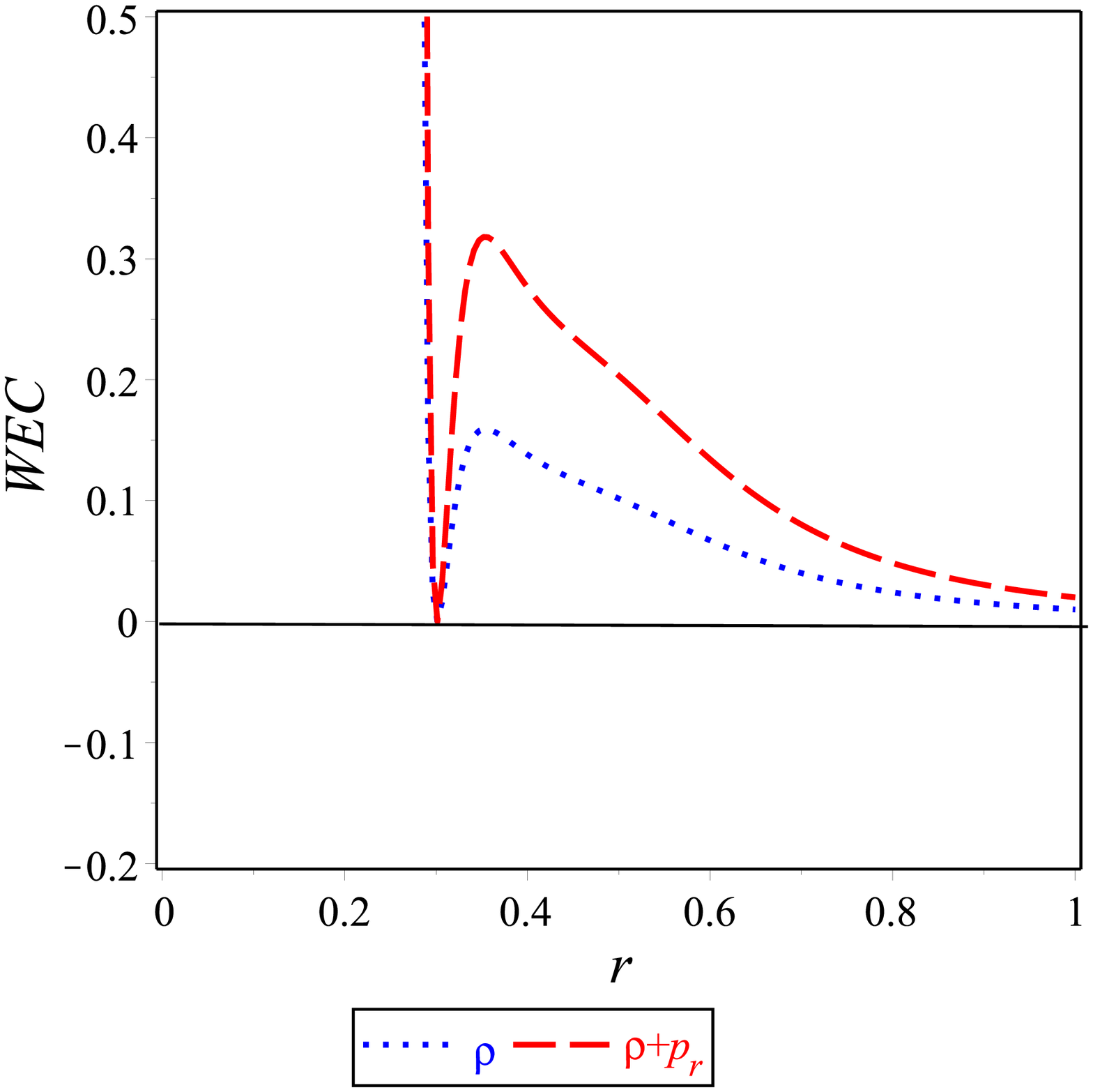}}\hspace{0.5cm}
\subfigure[~SEC]{\label{fig:2b}\includegraphics[scale=0.2]{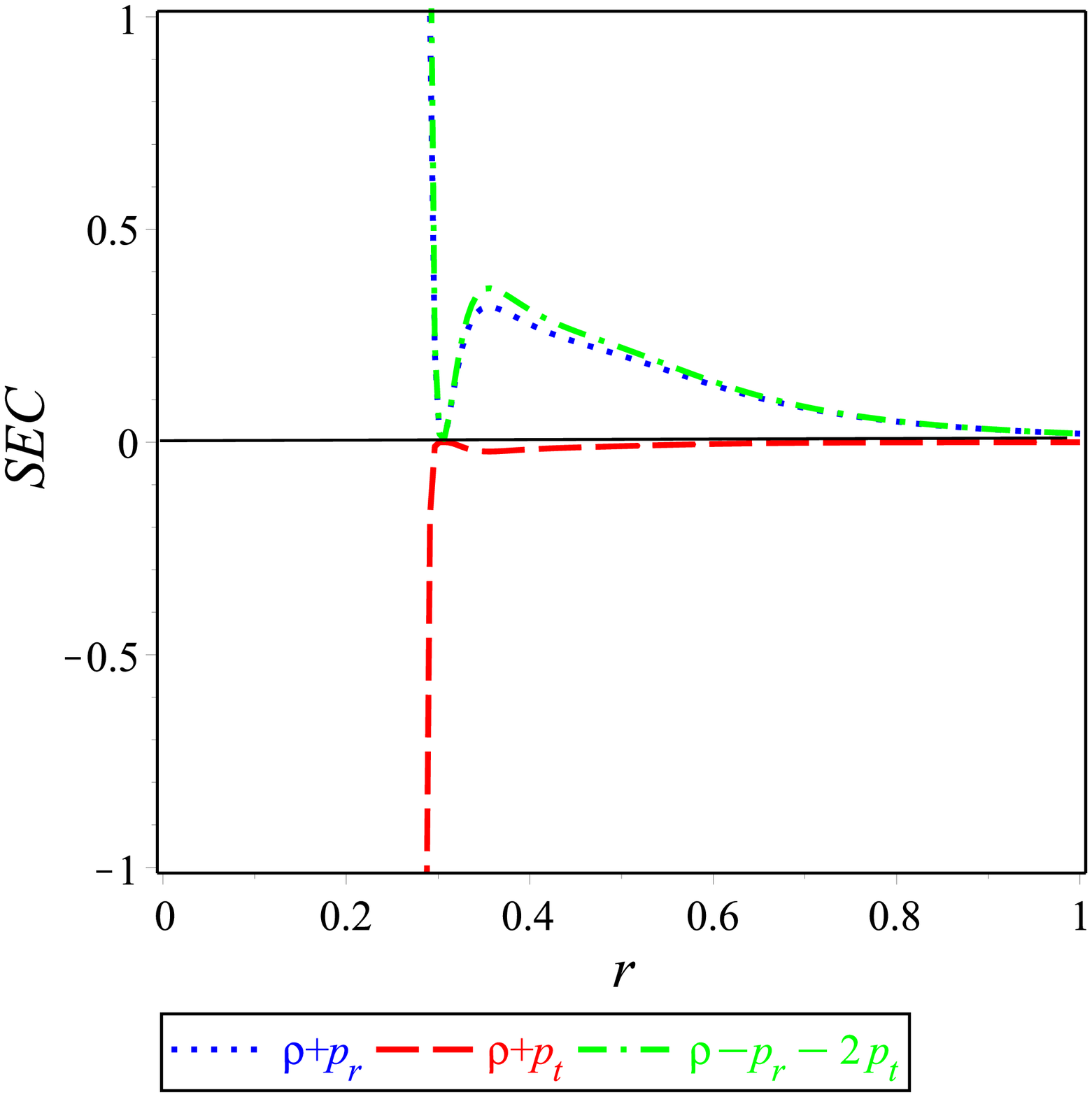}}\hspace{0.5cm}
\subfigure[~NEC]{\label{fig:2c}\includegraphics[scale=0.2]{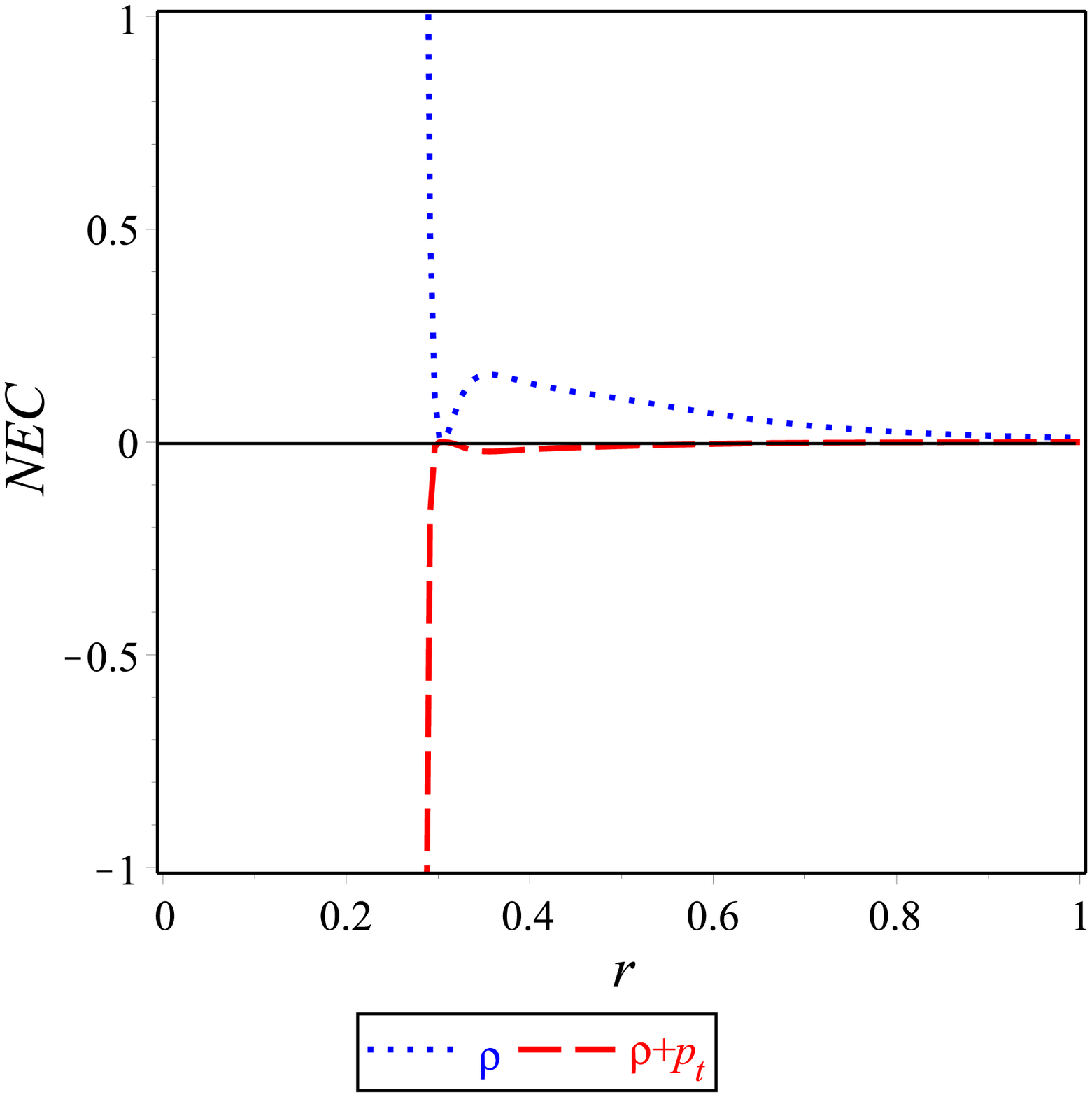}}\hspace{0.5cm}
\subfigure[~DEC]{\label{fig:2d}\includegraphics[scale=0.2]{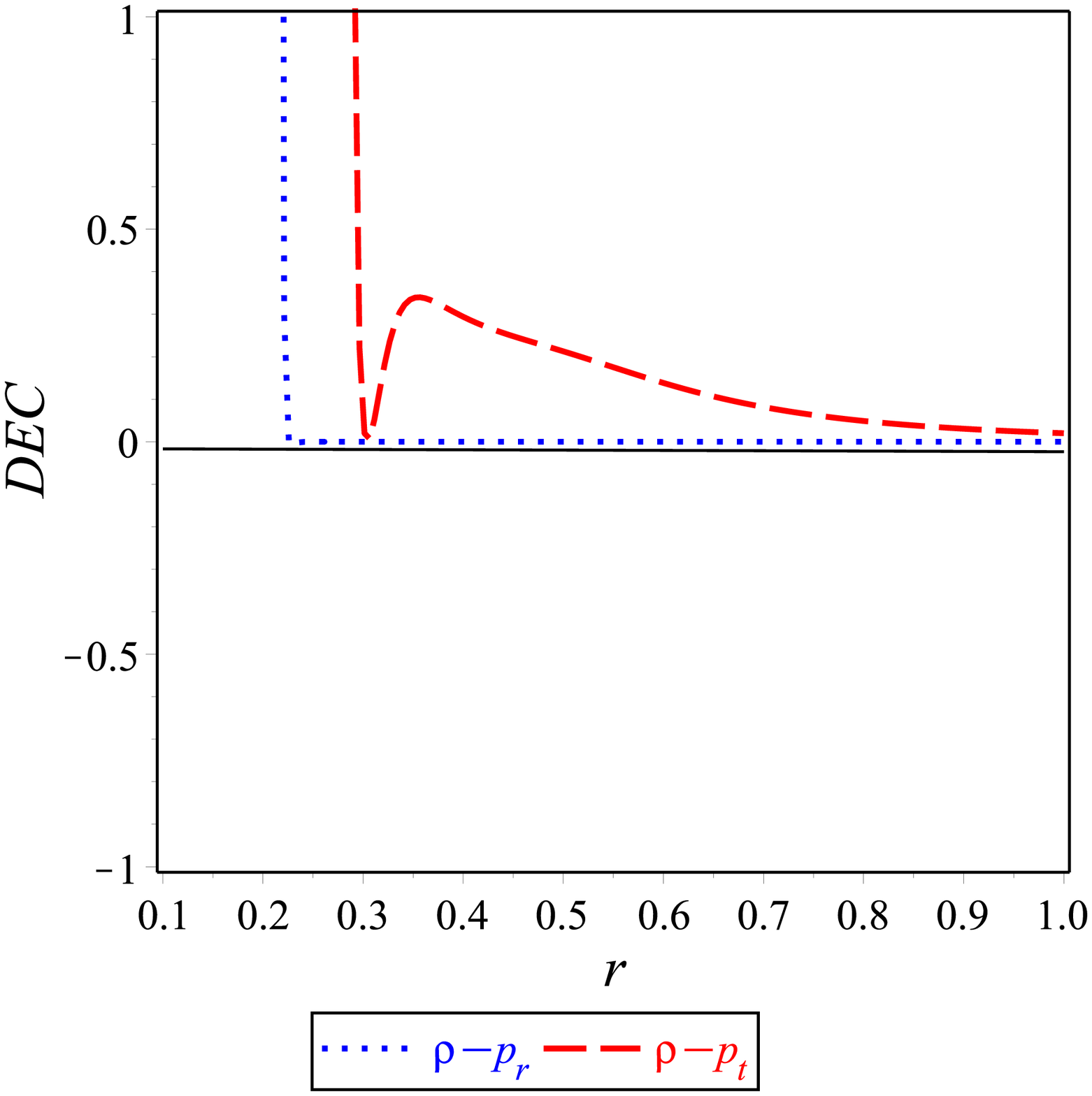}}
\caption{Schematic plots of the, WEC, SEC, NEC and DEC given by Eq.~(\ref{elm2}).}
\label{Fig:2}
\end{figure*}

\begin{figure*}
\centering
\subfigure[~WEC]{\label{fig:3a}\includegraphics[scale=0.2]{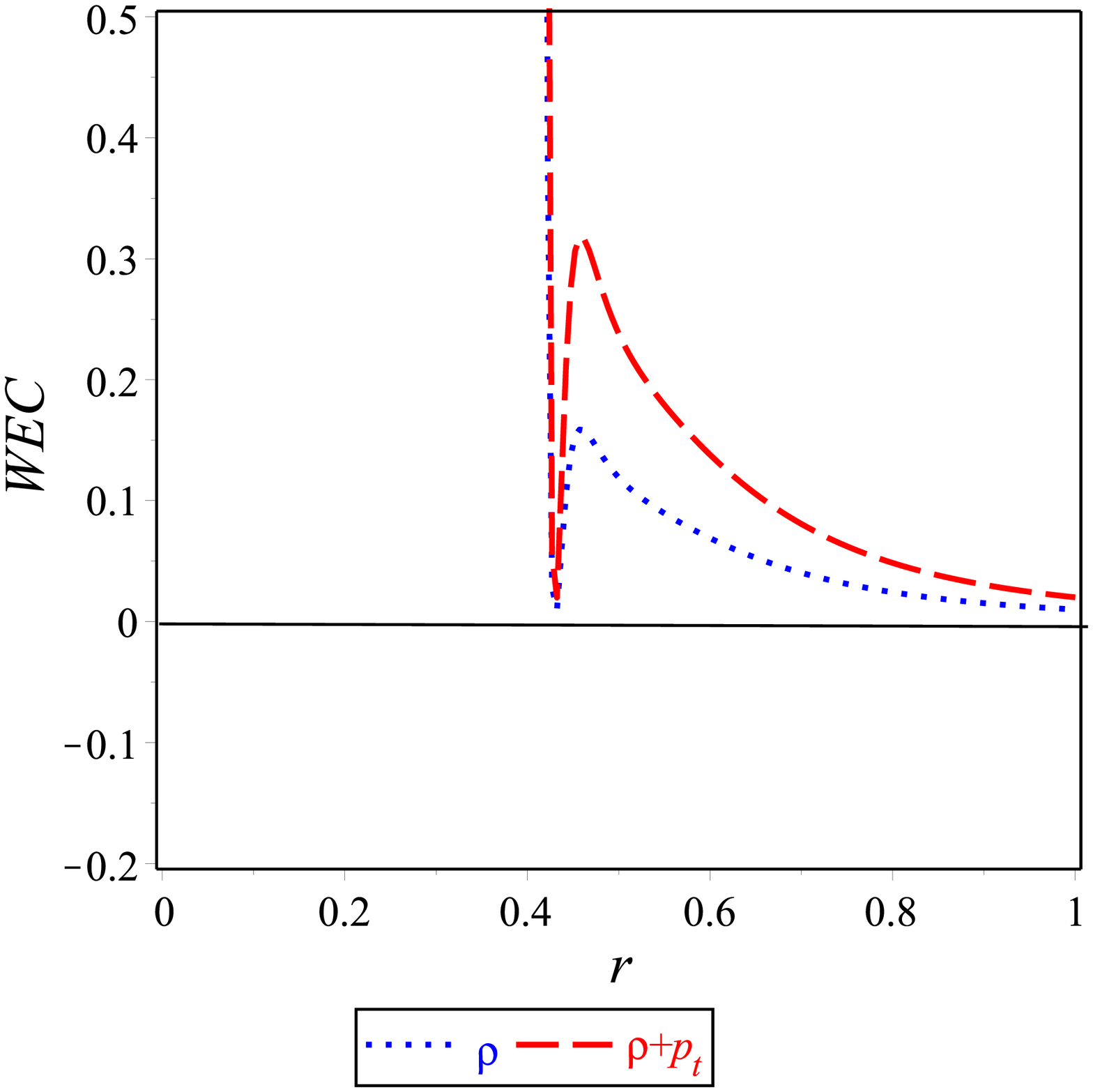}}\hspace{0.5cm}
\subfigure[~SEC]{\label{fig:3b}\includegraphics[scale=0.2]{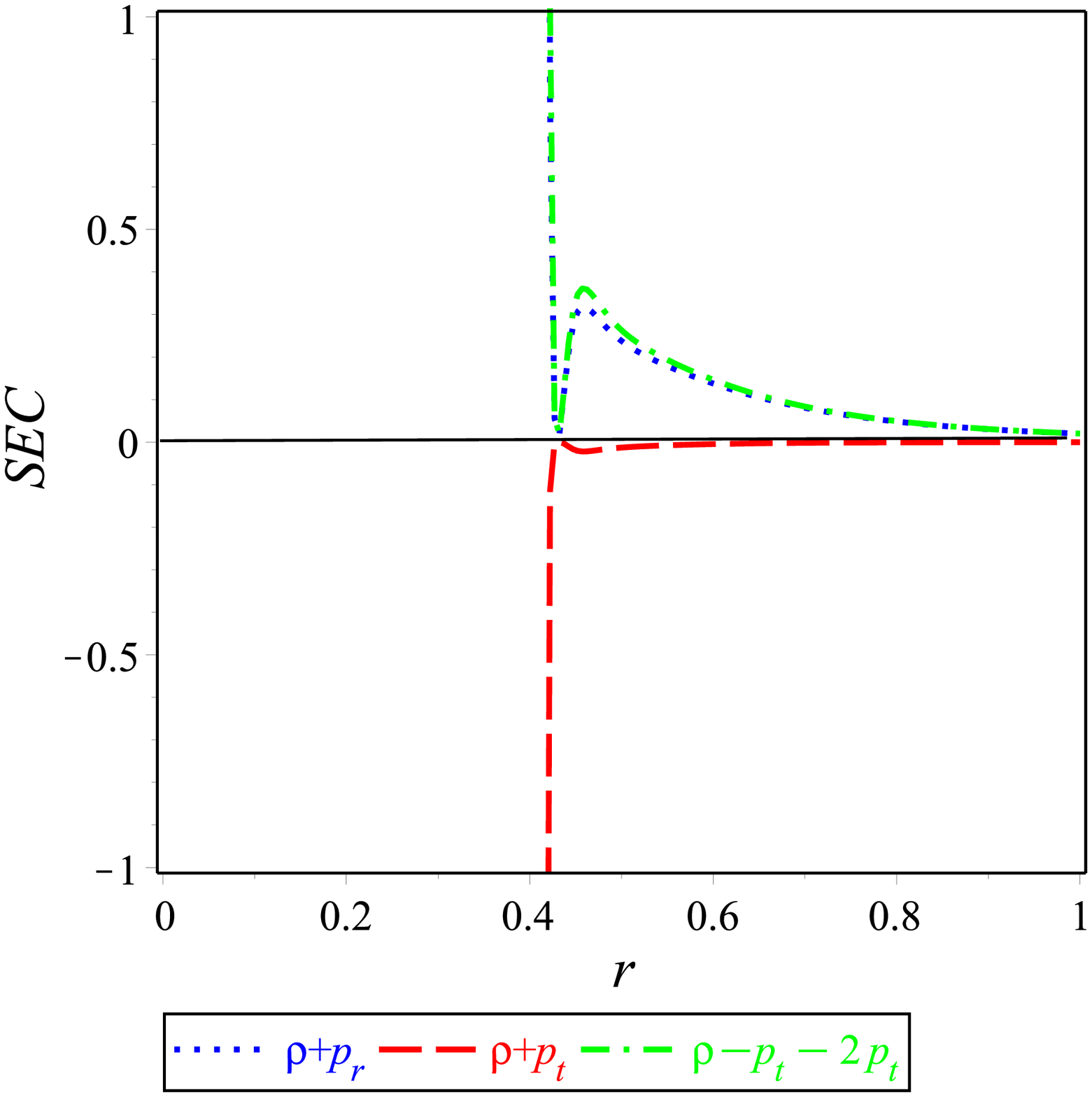}}\hspace{0.5cm}
\subfigure[~NEC]{\label{fig:3c}\includegraphics[scale=0.2]{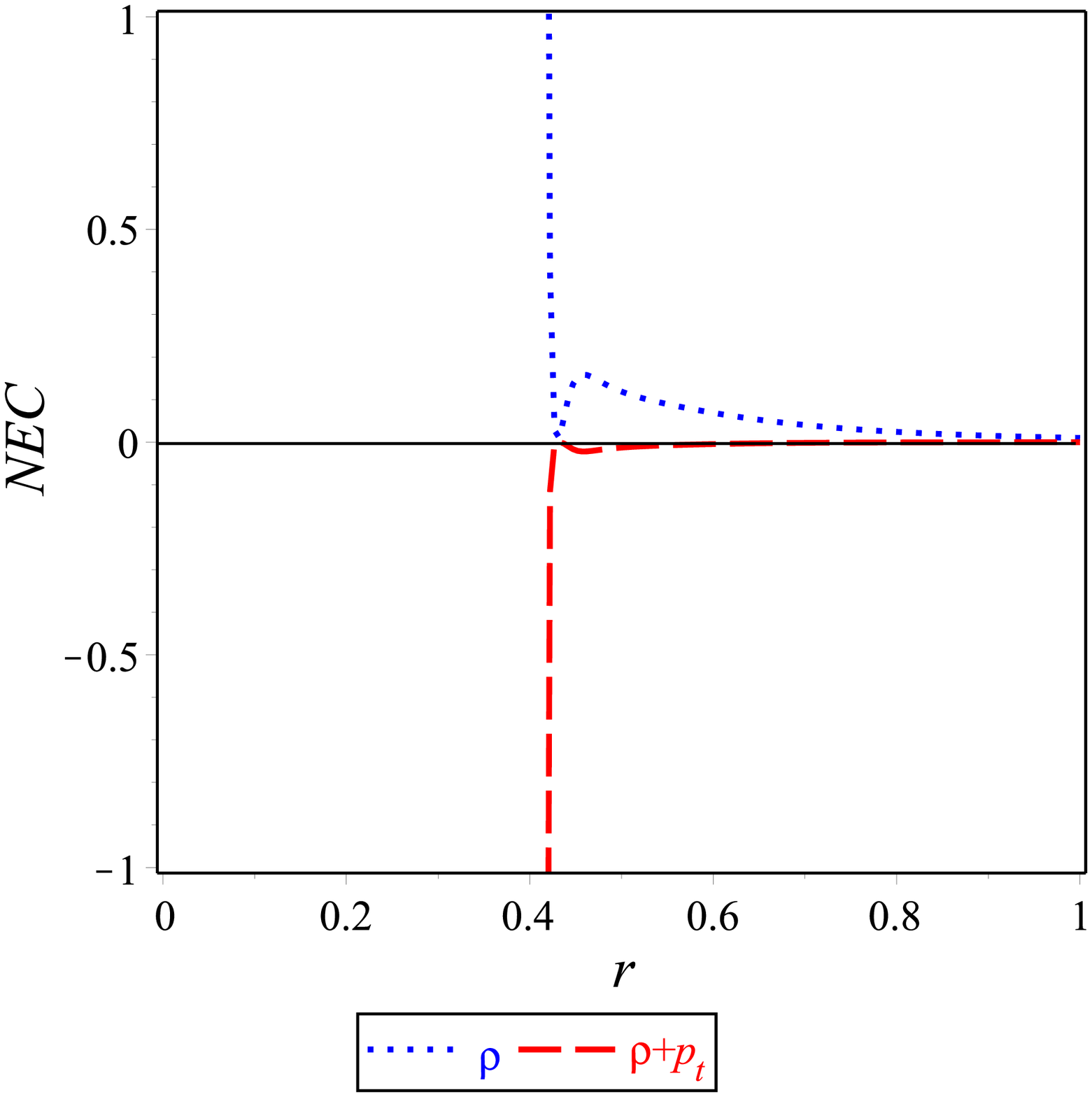}}\hspace{0.5cm}
\subfigure[~DEC]{\label{fig:3d}\includegraphics[scale=0.2]{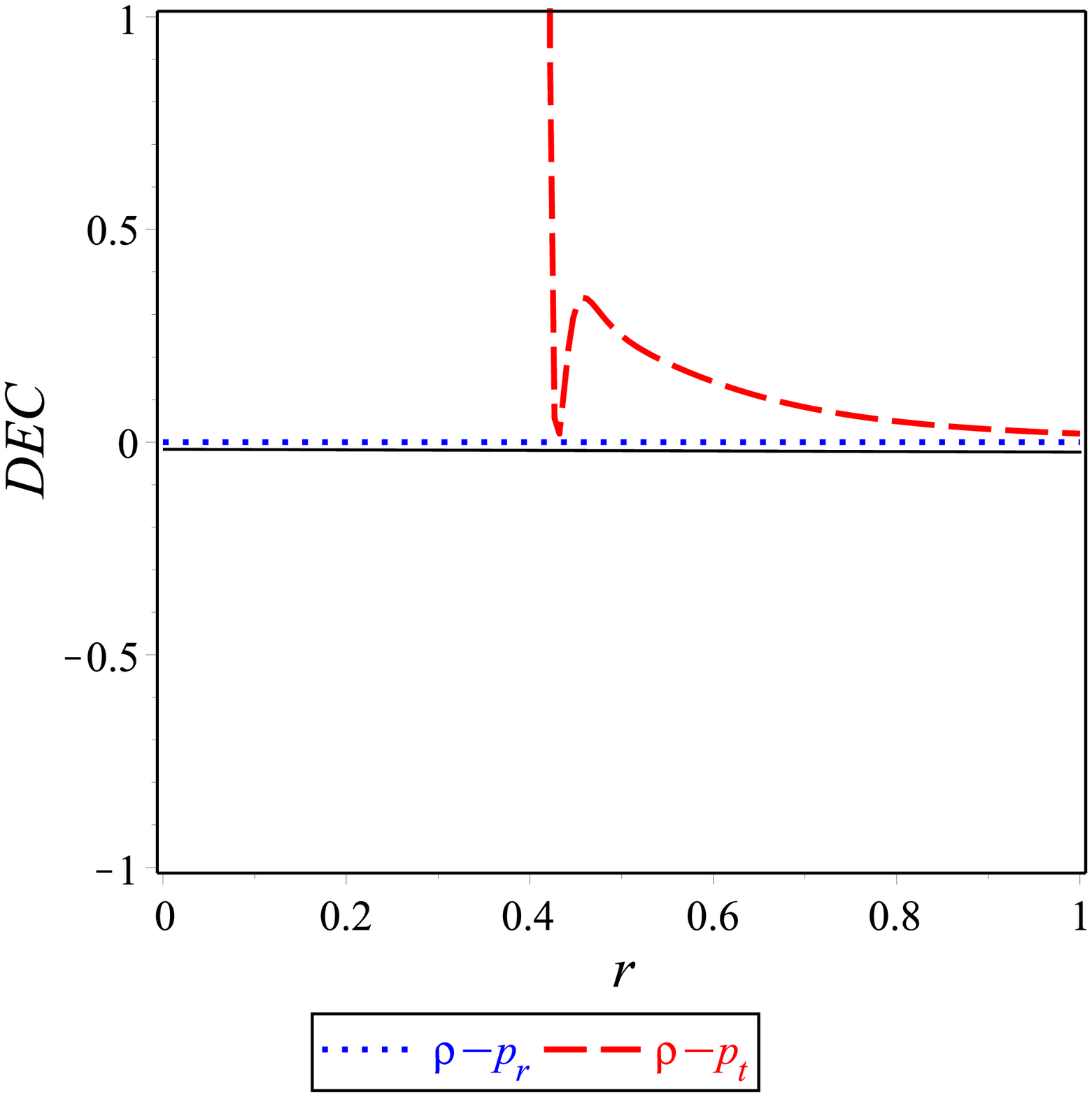}}
\caption{Schematic plots of WEC, SEC, NEC and DEC given by Eq.~(\ref{elm3}).}
\label{Fig:3}
\end{figure*}

Energy conditions provide important tools to examine and better understand cosmological models and/or strong gravitational fields.
We are interested in the study of energy conditions in the nonlinear electrodynamics case, because the linear case is well-known in GR theory.
As mentioned previously, we will focus on the solution of Case (\textbf{I}) of the nonlinear electrodynamic BH solution (\ref{sol1}).
The energy conditions are classified into four categories: strong energy (SEC), weak energy (WEC), null energy (NEC),
and dominant energy (DEC) conditions \cite{Hs73, Nashed:2016gwp}. To fulfill these conditions, the following inequalities must be satisfied:
\begin{eqnarray}
\label{ec}
\mbox{SEC}: & \quad \rho+p_r\geq 0\,, & \quad \rho+p_t\geq 0\,, \quad \rho-p_r-2p_t\geq 0\, , \nonumber\\
\mbox{WEC}: & \quad \rho\geq 0\,, & \quad \rho+p_r\geq 0\,, \nonumber\\
\mbox{NEC}: & \quad \rho \geq 0\, , & \quad \rho+p_t\geq 0\, , \nonumber\\
\mbox{DEC}: & \quad \rho\geq \left|p_r\right|\, , & \quad \rho\geq \left| p_t \right|\, ,
\end{eqnarray}
where ${\mathcal{T}^\mathrm{EH}}{_0}{^0}=\rho$, ${\mathcal{T}^\mathrm{EH}}{_1}{^1}=p_r$ and ${\mathcal{T}^\mathrm{EH}}{_2}{^2}={\mathcal{T}^\mathrm{EH}}{_3}{^3}=p_t$
are the density, radial, and tangential pressures, respectively.
Straightforward calculations of BH solution (\ref{elm1}) give
\begin{eqnarray}
\label{econ}
\mbox{SEC}: & \quad \rho+p_r\approx \frac{q^2(2r^4-q^2\mu)}{r^8}> 0\, , & \quad \rho+p_t\approx\frac{q^4\mu}{r^8}< 0\, , \quad
\rho-pr-2p_t\approx\frac{q^2(2r^4-3q^2\mu)}{r^8}> 0\, , \nonumber\\
\mbox{WEC}: & \quad \rho\approx \frac{q^2(2r^4-q^2\mu)}{2r^8}> 0\, , & \quad \rho+p_r= \frac{q^2(2r^4-q^2\mu)}{r^8}> 0\,, \nonumber\\
\mbox{NEC}: & \quad \rho \approx \frac{q^2(2r^4-q^2\mu)}{2r^8}> 0\, , & \quad \rho+p_t\approx\frac{q^4\mu}{r^8}< 0\,, \nonumber\\[10pt]
\mbox{DEC}: & \quad \rho\geq \left|p_r\right| \quad \mbox{(satisfied)}\, , & \quad \rho\geq \left| p_t \right| \quad \mbox{(satisfied)}\, .
\end{eqnarray}
This shows that SEC and NEC are not satisfied, which subsequently indicates that the solutions have ghost instabilities.
This is generally in agreement with previous studies on mimetic gravity.
It is worth mentioning that a suitable Lagrangian multiplier can be used to overcome the ghost instability in mimetic gravity \cite{Nojiri:2017ygt}.
This approach needs to be examined in the model at hand.
Remarkably, the SEC and NEC are not satisfied due to the contribution of parameter $\mu$, which characterizes the nonlinear electromagnetic charge.
This clearly shows how the nonlinear contribution of the charge strengthens the singularity, as discussed in the previous subsection.
The aforementioned discussion can be explained graphically, as indicated in Figure~\ref{Fig:1}, for the BH (\ref{elm1})
and the same procedures can be applied for BHs (\ref{elm2}) and (\ref{elm3}) plotted in Figures~\ref{Fig:2}~and~\ref{Fig:3}
Figure~\ref{Fig:1}~\subref{fig:1a}--\ref{Fig:1}~\subref{fig:1d} shows the energy conditions of solution (\ref{elm1}), which also coincides with Eq.~(\ref{econ}).
The main reason for the SEC and NEC breaking is the negative value of parameter $\mu$; however, if $\mu$ takes a positive value, WEC, SEC, and DEC are broken.

\section{Thermodynamics and stability }\label{S4}
We considered another physics approach to deeply elucidate the three BHs with (\ref{elm1}), (\ref{elm2}), and (\ref{elm3}) by investigating their thermodynamic behavior.
Accordingly, we will present the main tools of the thermodynamic quantities.

\subsection{Thermodynamics of the BH(\ref{elm1}) }\label{S5a}
\begin{figure*}
\centering
\subfigure[~Possible horizons]{\label{fig:4a}\includegraphics[scale=0.25]{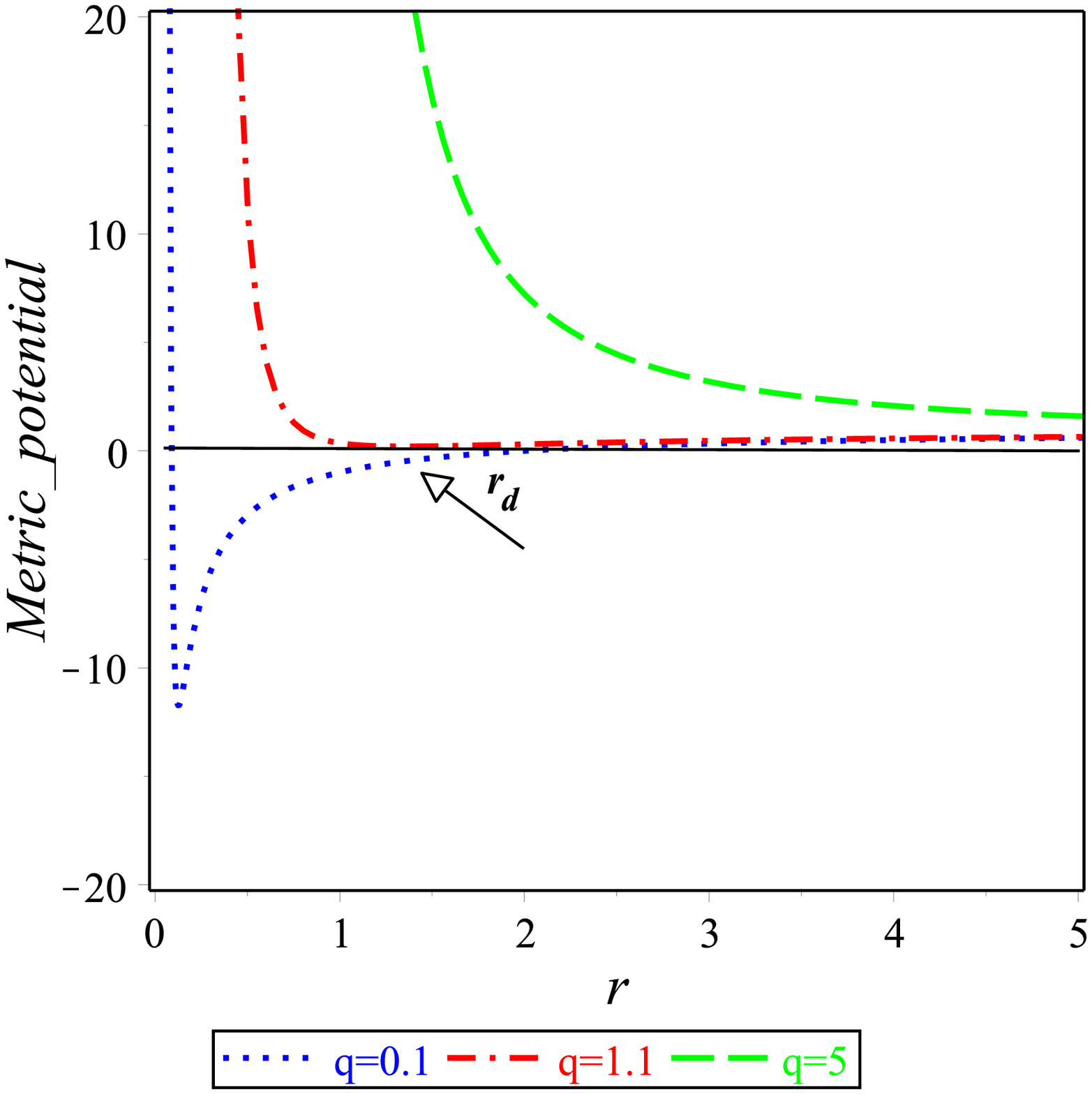}}\hspace{0.5cm}
\subfigure[~The horizon mass-radius]{\label{fig:4b}\includegraphics[scale=0.25]{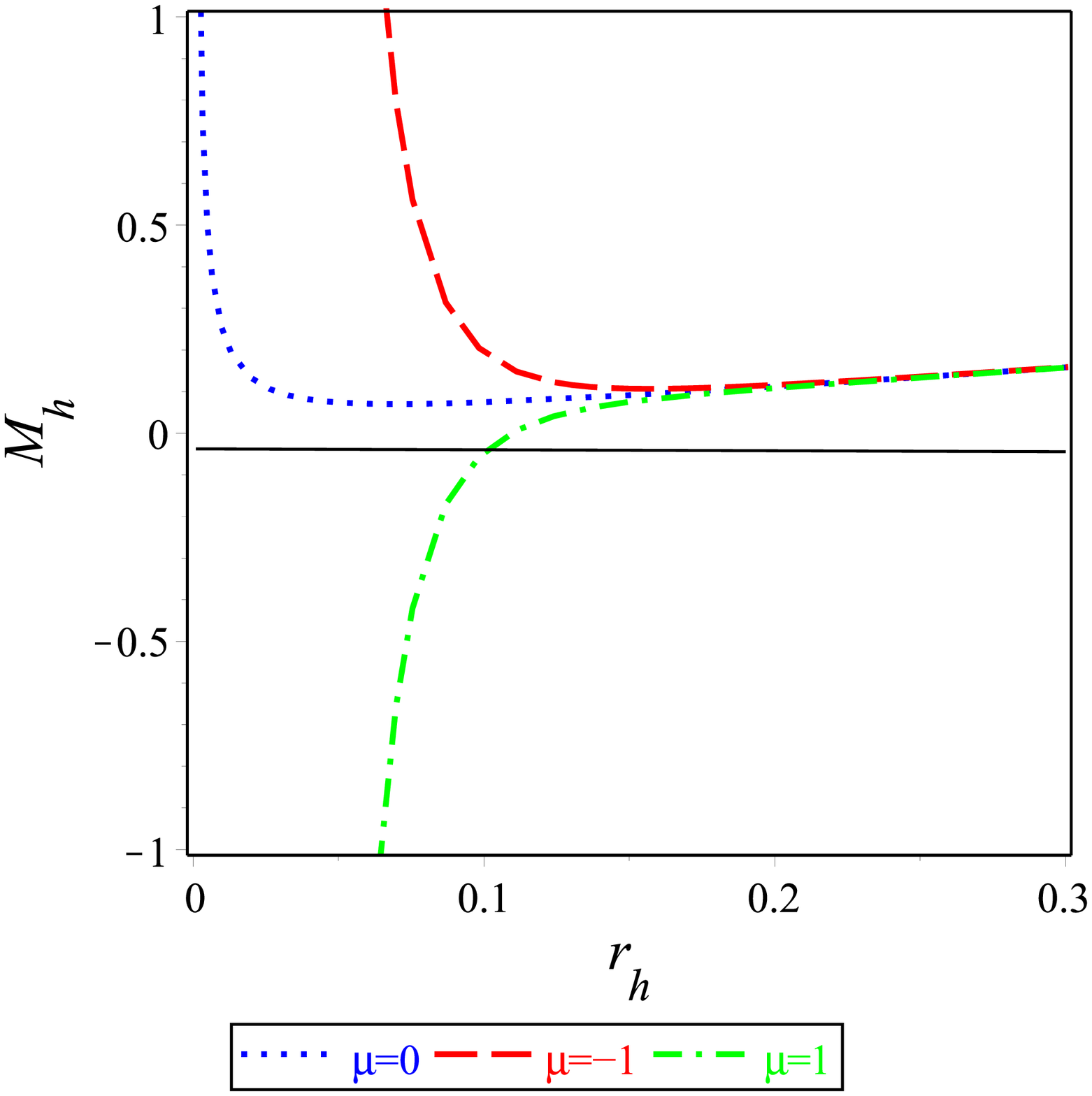}}\hspace{0.5cm}
\subfigure[~The horizon Bekenstein-Hawking entropy]{\label{fig:4c}\includegraphics[scale=0.25]{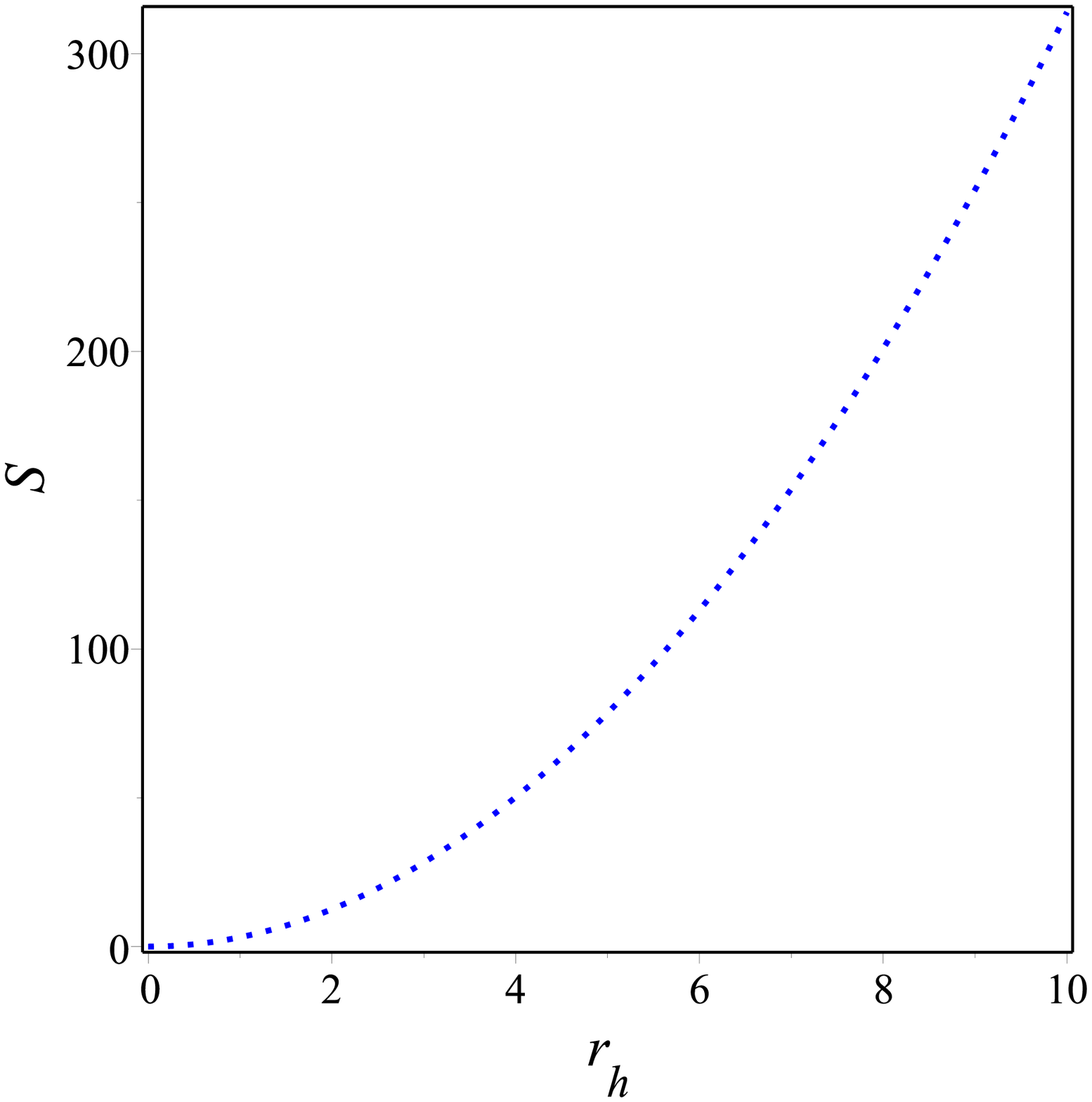}}\hspace{0.5cm}
\subfigure[~The horizon Hawking temperature]{\label{fig:4d}\includegraphics[scale=0.25]{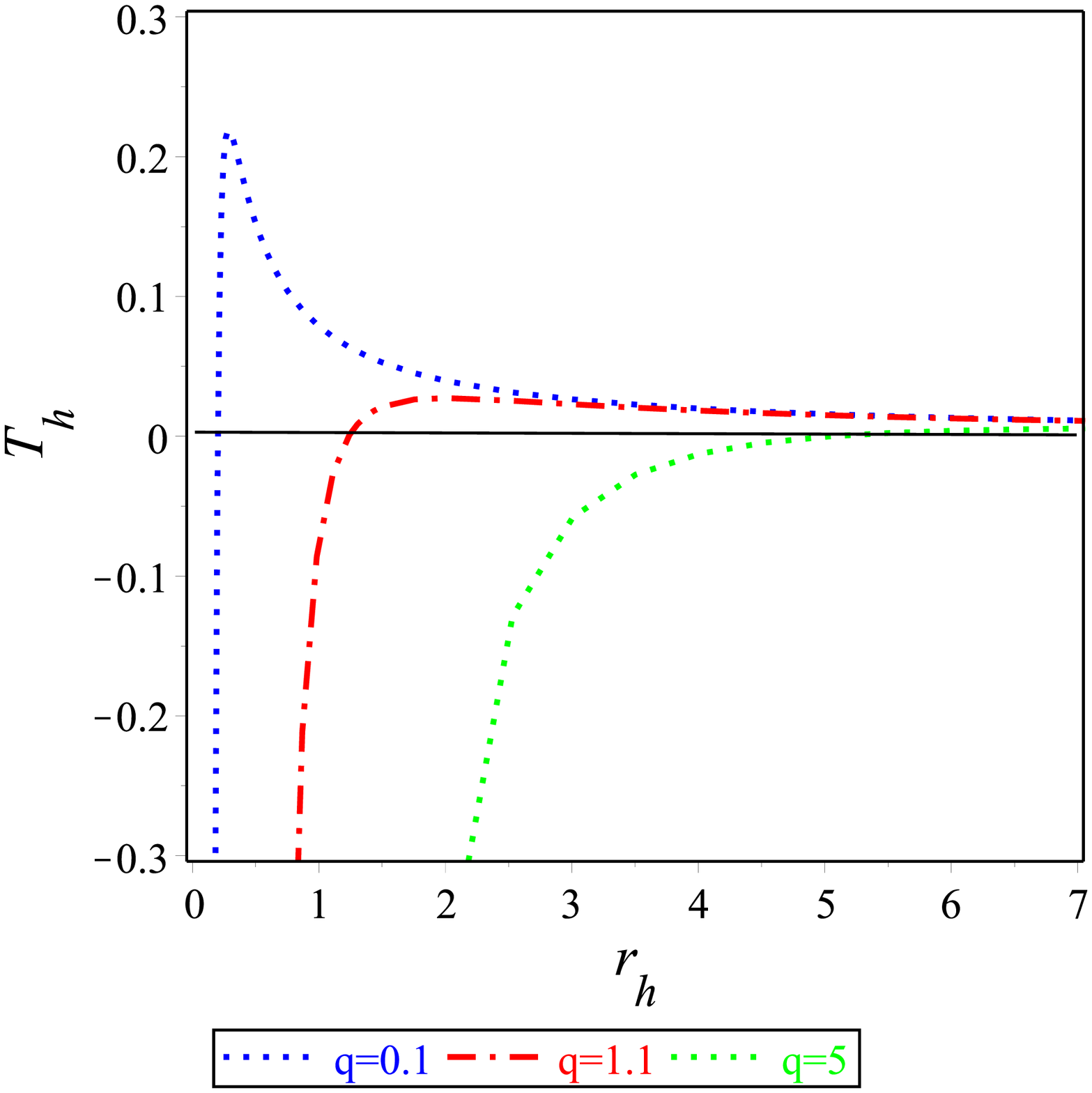}}\hspace{0.5cm}
\subfigure[~The horizon heat capacity]{\label{fig:4e}\includegraphics[scale=0.25]{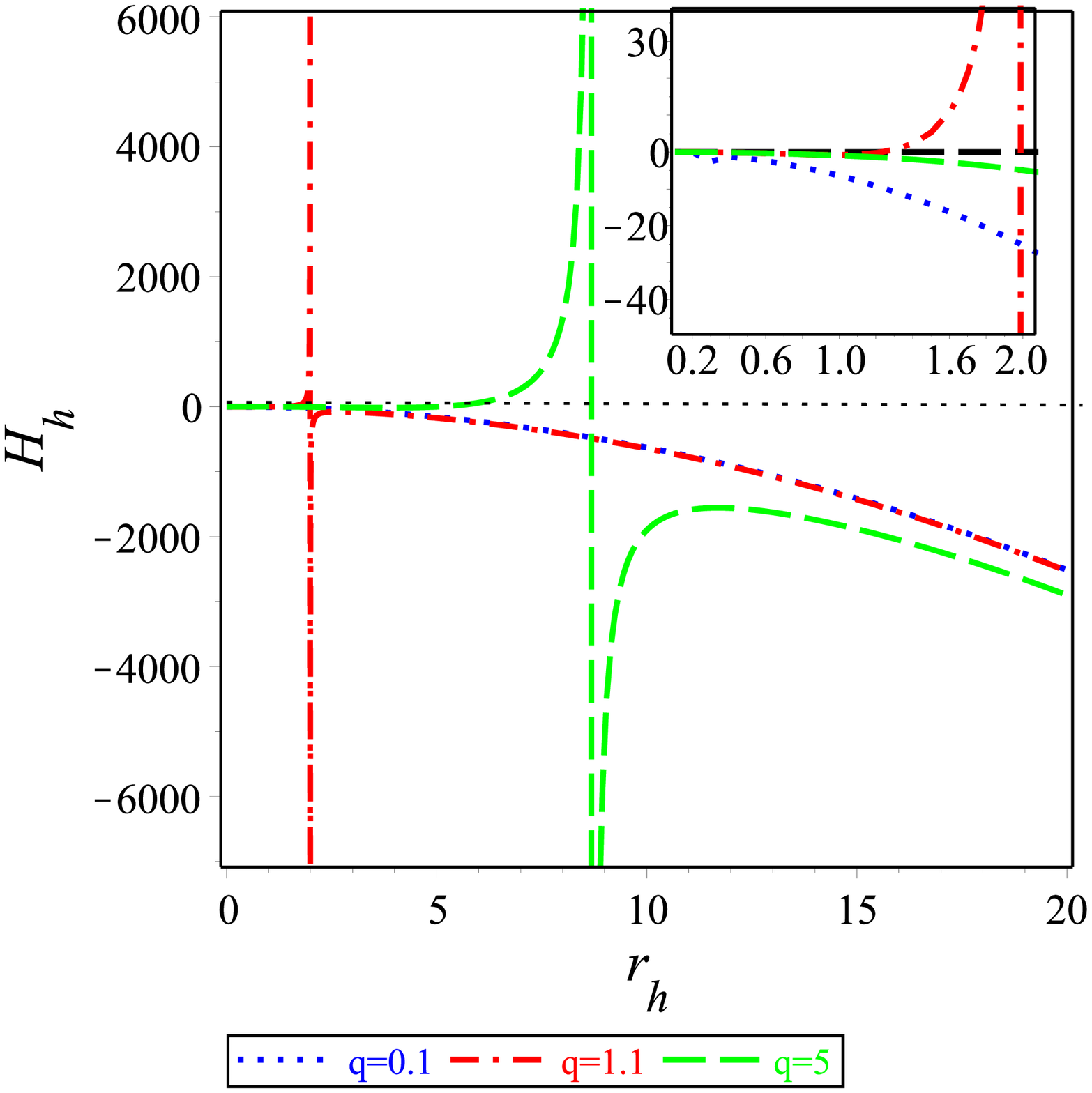}}\hspace{0.5cm}
\caption{Schematic plots of the thermodynamic quantities of the BH solution (\ref{elm1}) for the negative value of parameter $\mu$:~\subref{fig:4a}
typical behavior of metric function $f(r)$ obtained using Eq.~(\ref{hor11});~\subref{fig:4b} the mass--radius relation of the horizon (\ref{hor-mass-rad1a11});
\subref{fig:4c} typical behavior of the entropy of the horizon which indicates that $S_h$ increases quadratically as $r_h$ increases to always afford
a positive entropy;~\subref{fig:4d} typical behavior of the temperature of horizon (\ref{T11}), which indicates that we have negative values of $T_h$ when $r_h<r_d$
and when $r_h>r_d$, $T_h$ becomes positive and~\subref{fig:4e} the heat capacity, (\ref{heat-cap1a11}),
indicates that we obtain a positive heat capacity when $r_h<r_d$ which indicates that the BH is stable.
}
\label{Fig:4}
\end{figure*}

The metric potential of the temporal component of Eq.~(\ref{elm1}) is given as follows,
\begin{eqnarray}
\label{hor11}
&&f(r)=1-\frac{2M}{r}+\frac{q^2}{r^2}-\frac{q^4\mu}{10 r^6}\,.
\end{eqnarray}
The behavior of Eq.~(\ref{hor11}) is shown in Figure~\ref{Fig:4}~\subref{fig:4a}, which also indicates that the BH could possess two horizons at the root of $f(r)=0$.
These two horizons are $r_c$, which denotes the inner Cauchy horizon of the BH, and $r_h$, which is the outer event horizon.
Figure~\ref{Fig:4}~\subref{fig:4a} indicates that if parameter $\mu=-1$ and charge parameter $q=0.1$, we will obtain two horizons and when $q=1.1$,
the two horizons will coincide and form one horizon, i.e., the degenerate horizon, i.e., when $r_{c}=r_{h}=r_d$.
Finally, when $q=5$, we will enter a parameter region without a horizon parameter in which the central singularity will be a naked singularity.

The total mass contained within event horizon ($r_h$) can be calculated by setting $f(r_h) = 0$.
Afterward, the mass-radius relation of the horizon can be obtained as follows:
\begin{eqnarray}
\label{hor-mass-rad1a11}
&& {M_h}=\frac{10r_h{}^4 \left( r_h{}^2+q^2\right)-q^4\mu}{20r_h{}^5}\, .
\end{eqnarray}
To present the aforementioned features differently, Figure~\ref{Fig:4}~\subref{fig:4b} depicts
value $M_h$, which corresponds to the $r_h$ horizon.
As Figure~\ref{Fig:4}~\subref{fig:4b} shows that for $\mu=0$ or $\mu=-1$, there is no root for Eq.~(\ref{hor-mass-rad1a11}) while for $\mu=1$ there is one root.

The Hawking temperature is generally defined as follows \cite{PhysRevD.86.024013,Sheykhi:2010zz,Hendi:2010gq,PhysRevD.81.084040}
\begin{equation}
\label{temp11}
T_h = \frac{f'(r_h)}{4\pi}\, ,
\end{equation}
where event horizon $r = r_h$ is the positive solution of equation $f(r_h) = 0$, which satisfies $f'(r_h)\neq 0$.
The entropy is given by \cite{PhysRevD.84.023515,Zheng:2018fyn}
\begin{equation}
\label{ent11}
S(r_h)=\frac{1}{4}A\, ,
\end{equation}
where $A$ is the area of the horizon.
The constraint, $f(r_h) = 0$, yields a six order algebraic equation and at $\mu=0$, we obtain the GR limit, $r_h=M\pm\sqrt{M^2-q^2}$.
 From Eq.~(\ref{ent11}), the entropy of Eq.~(\ref{hor11}) assumes the following form:
\begin{eqnarray}
\label{ent1}
{S_h}
=\pi r_h{}^2\,,
\end{eqnarray}
which indicates that the entropy does not affect by the nonlinear expression of electrodynamics.
The behavior of Eq.~(\ref{ent1}) is shown in Figure~\ref{Fig:4}~\subref{fig:4c} which expresses a positive entropy value.

Using Eq.~(\ref{temp11}), the Hawking temperature can be calculated as:
\begin{eqnarray}
\label{T11}
&&T_h=\frac{2r_h{}^4 \left( r_h{}^2-q^2 \right)+q^4\mu}{8\pi\,r_h{}^7}\, .
\end{eqnarray}
The behavior of the Hawking temperature given by Eq.~(\ref{T11}) is drawn in Figure~\ref{Fig:4}~\subref{fig:4d} which indicates that $T_h$ has a vanishing value
at $r_h = r_d$ for different values of charge parameter $q$.
Moreover, when $r_h < r_d$, $T_h$ becomes negative and an ultracold BH is formed.
Additionally, Davies \cite{Davies:1978mf} clarified that there is no clear reason for thermodynamical effects to prevent the BH temperature from being below the absolute zero;
in that case, a naked singularity is formed.
Figure~\ref{Fig:4}~\subref{fig:4d} shows Davies' argument at the $r_h < r_{d}$ region.

Furthermore, the stability of the BH solution is an essential topic that can be studied at the dynamic and perturbative levels
\cite{Nashed:2003ee,Myung:2011we,Myung:2013oca}.
To investigate the thermodynamic stability of BHs, the formula of the heat capacity $H(r_h)$
at the event horizon must be derived. It is defined as follows \cite{Nouicer:2007pu,DK11,Chamblin:1999tk}:
\begin{equation}
\label{heat-capacity11}
H_h\equiv H(r_h)=\frac{\partial M_h}{\partial T_h}=\frac{\partial M_h}{\partial r_h} \left(\frac{\partial T_h}{\partial r_h}\right)^{-1}\, .
\end{equation}
The BH will be thermodynamically stable, if its heat capacity $H_h$ is positive.
However, it will be unstable if $H_h$ is negative.
Substituting (\ref{hor-mass-rad1a11}) and (\ref{T11}) into (\ref{heat-capacity11}), we obtain the heat capacity as follows:
\begin{equation}
\label{heat-cap1a11}
{H_h}
=\frac{2\pi r_h{}^2 \left[2r_h{}^4 \left(r_h{}^2-q^2 \right) +q^4\mu\right]}{2r_h{}^4 \left( 3q^2-r_h{}^2 \right)-7q^4\mu} \, .
\end{equation}
Equation~(\ref{heat-cap1a11}) shows that $H_h$ does not locally diverge and that the BH exhibits a phase transition of the second order.
The heat capacity is depicted in Figure~\ref{Fig:4}~\subref{fig:4e} which also shows that $H_h<0$ when $\mu=-1$ for different values of charge parameter $q$.
The heat capacity is negative primarily because of the derivative of the Hawking temperature consistent with the nature of MEH and
the Reissner Nordstr\"om BHs which can be discovered at $\mu=0$.
It is important to note that we have a positive heat capacity when $r_h<r_d$ otherwise, we have a negative value.


\subsection{Thermodynamics of the BH (\ref{elm2}) }\label{S5b}
\begin{figure*}
\centering
\subfigure[~Possible horizons]{\label{fig:5a}\includegraphics[scale=0.25]{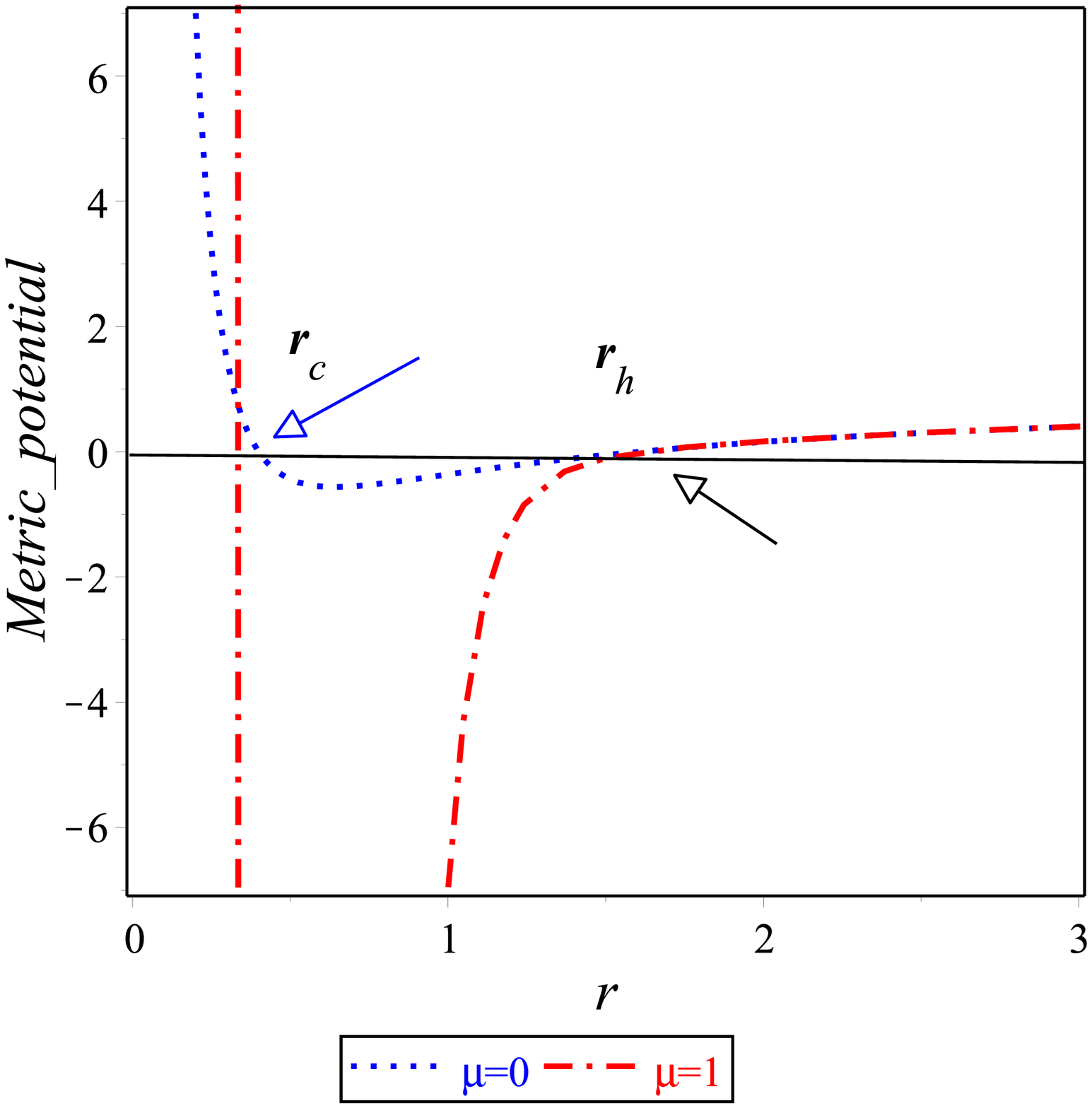}}\hspace{0.5cm}
\subfigure[~The horizon mass-radius]{\label{fig:5b}\includegraphics[scale=0.25]{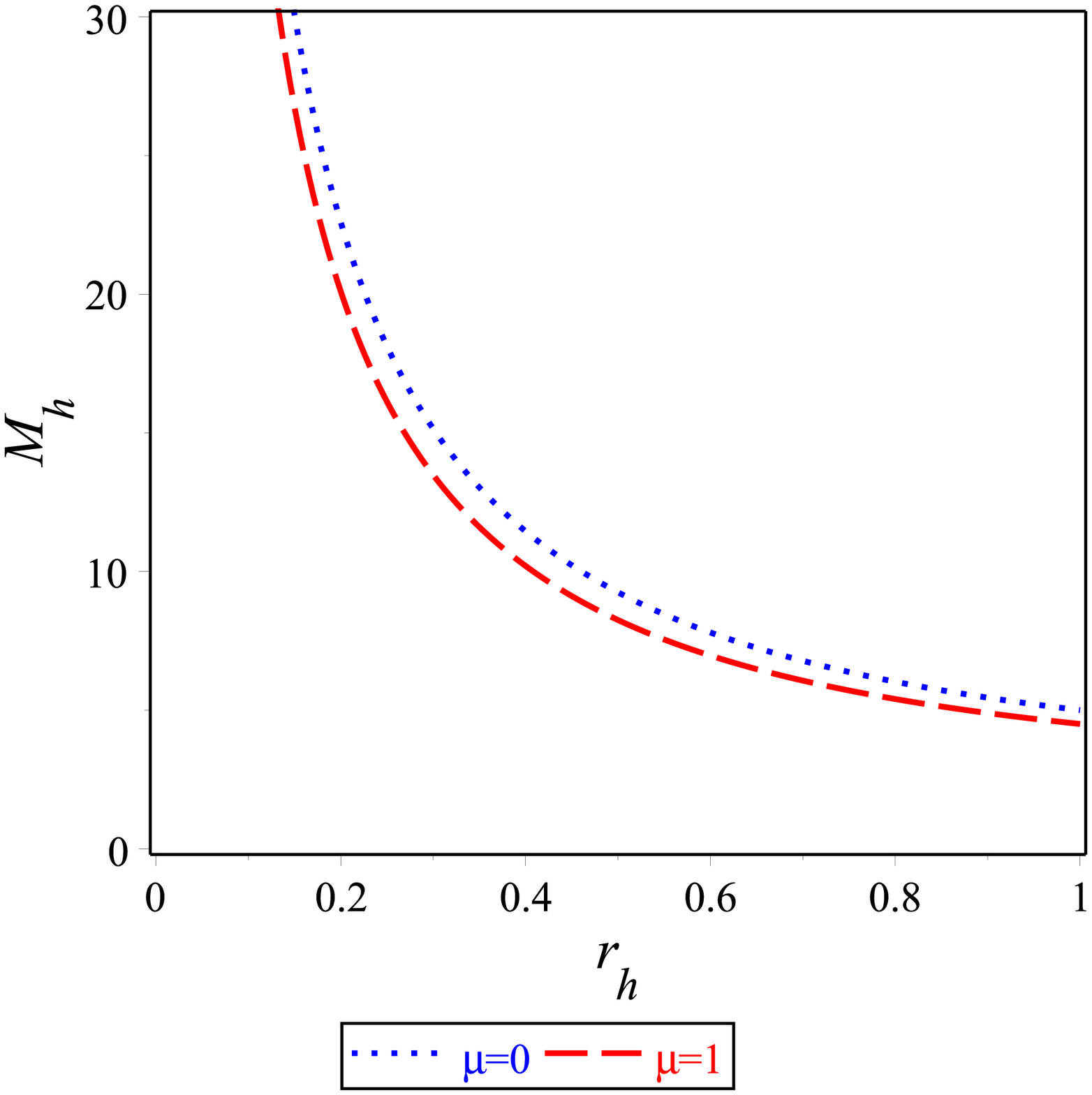}}\hspace{0.5cm}
\subfigure[~The horizon Hawking Temperature]{\label{fig:5c}\includegraphics[scale=0.25]{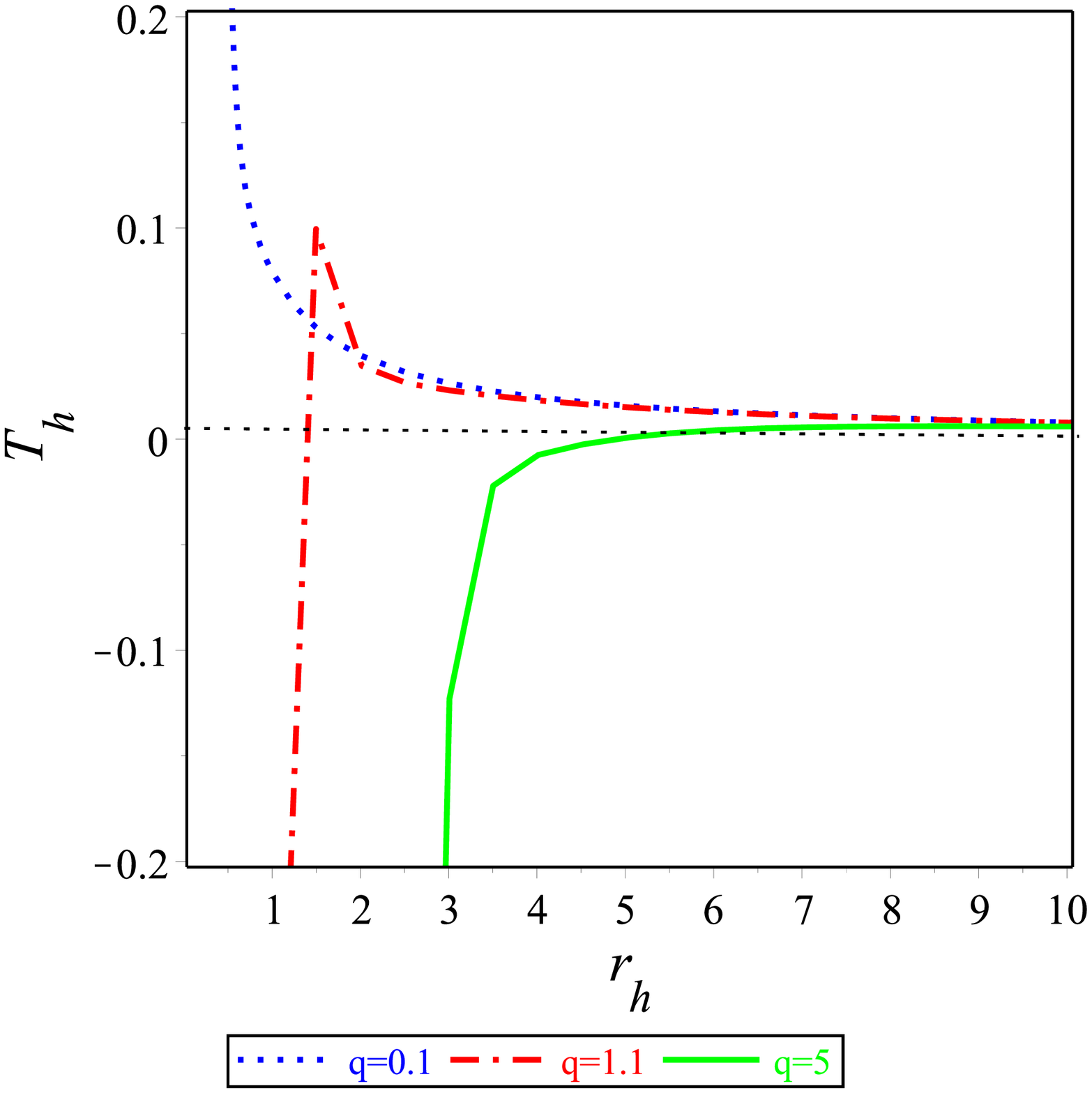}}\hspace{0.5cm}
\subfigure[~The horizon heat capacity]{\label{fig:5d}\includegraphics[scale=0.25]{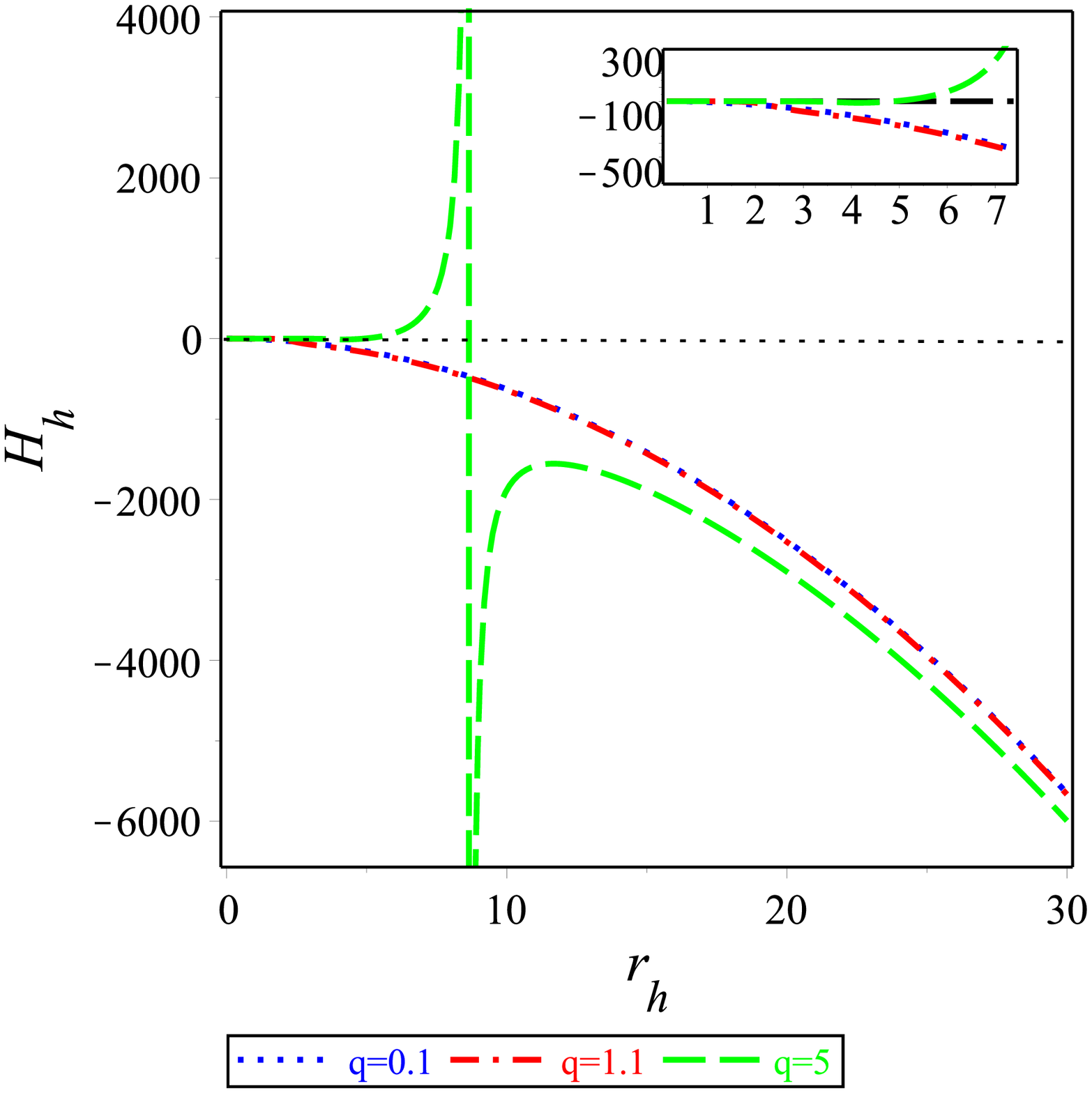}}\hspace{0.5cm}
\caption{Schematic plots of the thermodynamic quantities of the BH solution (\ref{elm2}) for the negative value of parameter $\mu$:~\subref{fig:5a}
typical behavior of metric function $f(r)$ obtained using (\ref{hor11});~\subref{fig:5b} the mass--radius relation of
horizon (\ref{hor-mass-rad1a22});~\subref{fig:4c} typical behavior of the temperature of horizon (\ref{T22}) which indicates
that the negative value happens when { $r_h>r_d$} and that { $r_h<r_d$} would becomes positive and~\subref{fig:4d} the heat capacity, (\ref{heat-cap1a22}),
indicates that we obtain a negative heat capacity.
}
\label{Fig:5}
\end{figure*}

The metric potential of the temporal component of Eq.~(\ref{elm2}) is given by
\begin{eqnarray}
\label{hor22}
&&f(r)=1-\frac{2M}{r}+\frac{q^2}{r^2}-\frac{q^4\mu}{10 r^6}+\frac{36q^4\mu^2}{ r^8}-\frac{72Mq^4\mu^2}{ r^9}+\frac{32q^6\mu^2}{ r^{10}}\,,
\end{eqnarray}
Equation~(\ref{hor22}) is shown in Figure~\ref{Fig:5}~\subref{fig:5a}, which also shows that the BH (\ref{hor22}) could possess two horizons at the root of $f(r)=0$.
The same discussions conducted for the BH (\ref{elm1}) are also valid for the BH (\ref{elm2}).

The total mass contained within the event horizon ($r_h$) can be calculated by setting $f(r_h) = 0$.
Afterward, the mass-radius of the horizon can be obtained as follows:
\begin{eqnarray}
\label{hor-mass-rad1a22}
&& {M_h}=\frac{10r_h{}^8 \left( r_h{}^2+q^2 \right)-q^4\mu \left[ r_h{}^2 \left( r_h{}^2-360\mu \right) -320q^2\mu \right]}{20r_h \left( r_h{}^8+36q^4\mu^2 \right)}\, .
\end{eqnarray}
Using Eq.~(\ref{temp11}), the Hawking temperature of BH (\ref{elm2}) can be calculated as:
\begin{eqnarray}
\label{T22}
{T_h}=\frac{10r_h{}^{16} \left( r_h{}^2-q^2 \right) + q^4\mu \left[ r_h{}^2 \left\{5r_h{}^8 \left(r_h{}^2+144\mu \right)-360r_h{}^6q^2\mu-108\mu^2q^4
\left( r_h{}^2-120\mu \right) \right\}-11520q^6\mu^3 \right]}{10r_h{}^{11} \left( r_h{}^8+36q^4\mu^2 \right)}\,,
\end{eqnarray}
The behavior of the Hawking temperature given by Eq.~(\ref{T22}) is displayed in Figure~\ref{Fig:5}~\subref{fig:5c}
which indicates that $T_h$ has a vanishing value at $r_h = r_d$ for different values of charge parameter $q$.
The same discussions conducted for BH (\ref{hor11}) can also be followed for the BH (\ref{hor22}).

Substituting (\ref{hor-mass-rad1a22}) and (\ref{T22}) into (\ref{heat-capacity11}), we obtain the heat capacity as follows:
\begin{align}
\label{heat-cap1a22}
{H_h}
=&\left\{2\pi r_h{}^{10} \left[10r_h{}^{16} \left( r_h{}^2-q^2 \right) +q^4\mu \left[r_h{}^2 \left\{ 5r_h{}^8 \left( r_h{}^2+144\mu \right)
 -360r_h{}^6q^2\mu-108\mu^2q^4 \left( r_h{}^2-120\mu \right) \right\}-11520q^6\mu^3 \right] \right] \right\}\nonumber\\
&
\times \left\{ 1800r^{12}q^8\mu^3+27216r^4q^{12}\mu^5-246240r^{10}q^8\mu^4-4199040r^2q^{12}\mu^6+2160r^{16}q^6\mu^2+257760r^8q^{10}\mu^4-35r^{20}q^4\mu \right. \nonumber\\
& \left. -3960r^{18} q^4\mu^2+30r^{24}q^2-10r^{26}+4561920q^{14}\mu^6 \right\}^{-1} \,.
\end{align}
Equation~(\ref{heat-cap1a22}) shows that $H_h$ does not locally diverge and that the BH exhibits a phase transition of the second-order.
The heat capacity is depicted in Figure~\ref{Fig:5}~\subref{fig:5d} which also shows that $H_h<0$ when $\mu=-1$ for different values of charge parameter $q$.
The heat capacity is negative primarily because of the derivative of Hawking temperature is consistent with the nature of the MEH and Reissner Nordstr\"om BHs
which can be discovered at $\mu=0$.


\subsection{Thermodynamics of the BH (\ref{elm3}) }\label{S5c}
\begin{figure*}
\centering
\subfigure[~Possible horizons]{\label{fig:6a}\includegraphics[scale=0.25]{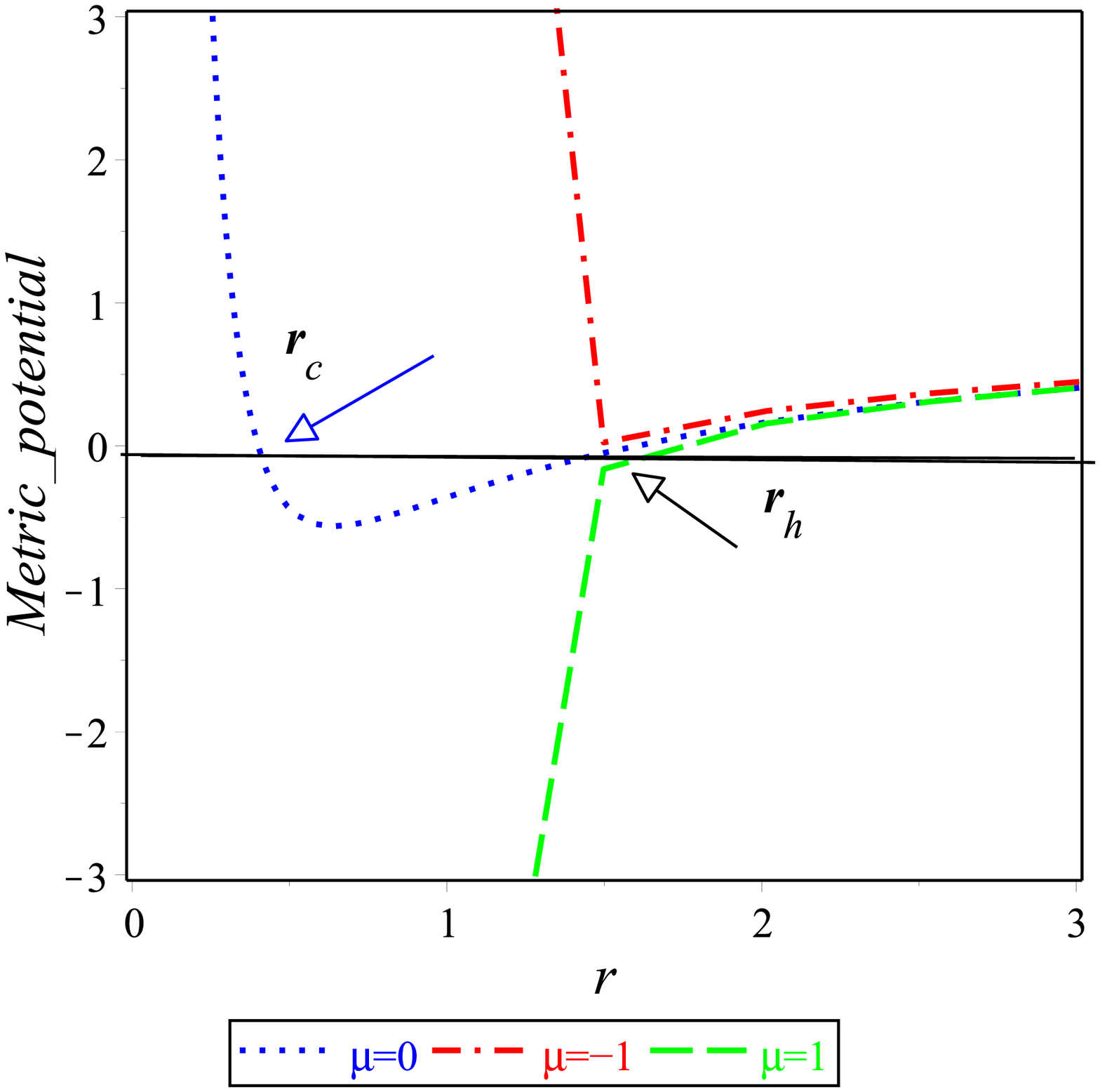}}\hspace{0.5cm}
\subfigure[~The horizon mass-radius]{\label{fig:6b}\includegraphics[scale=0.25]{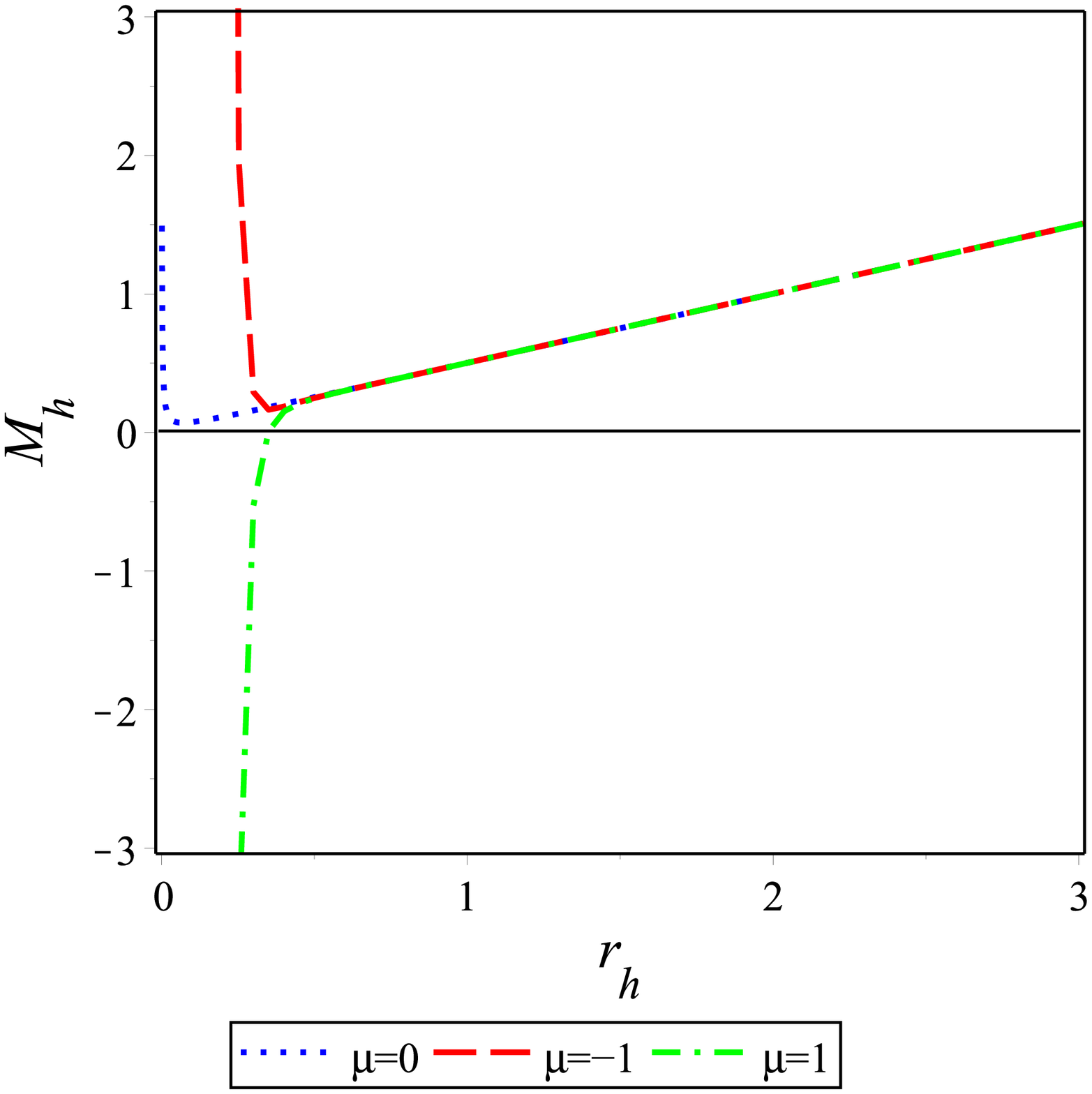}}\hspace{0.5cm}
\subfigure[~The horizon Hawking Temperature]{\label{fig:6c}\includegraphics[scale=0.25]{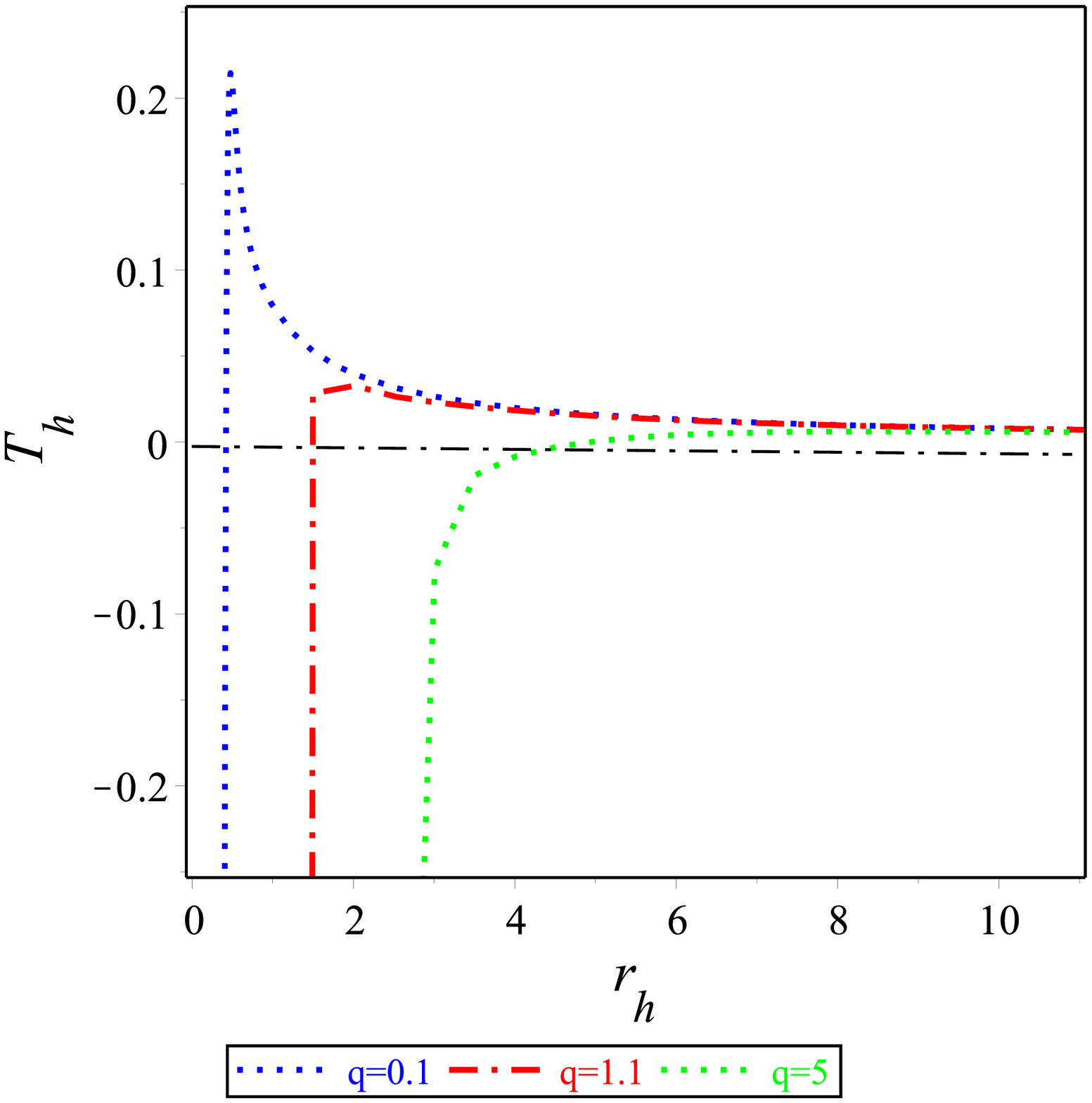}}\hspace{0.5cm}
\subfigure[~The horizon heat capacity]{\label{fig:6d}\includegraphics[scale=0.25]{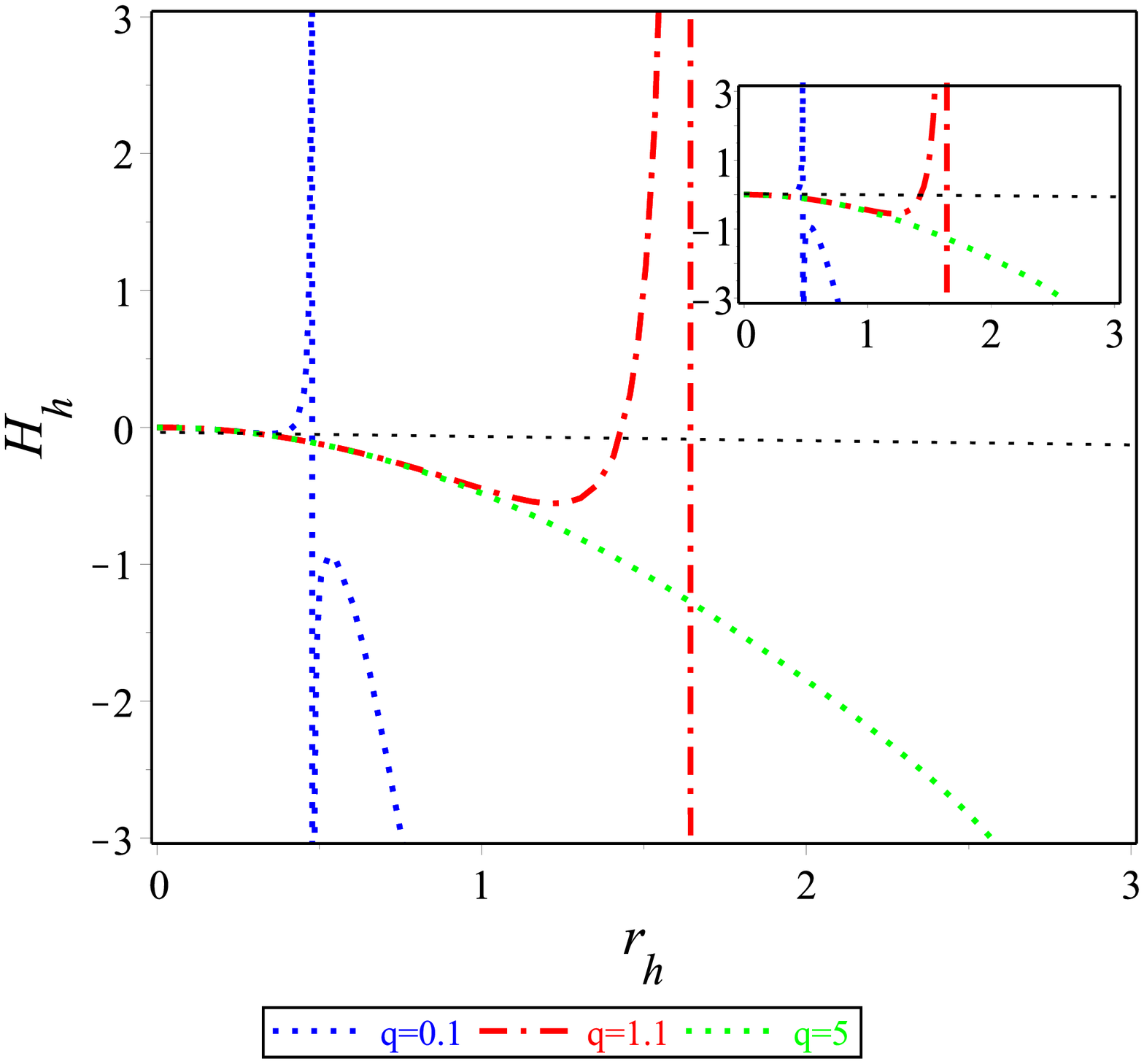}}\hspace{0.5cm}
\caption{Schematic plots of the thermodynamic quantities of the BH solution (\ref{elm3}) for the negative value of parameter $\mu$:~\subref{fig:6a}
typical behavior of the metric function $f(r)$ that was obtained using (\ref{hor33});~\subref{fig:6b} the mass--radius relation of the horizon (\ref{hor-mass-rad1a33});~\subref{fig:6c}
typical behavior of the temperature of the horizon (\ref{T33}) which indicates a negative value when { $r_h>r_d$} and {$r_h<r_d$} would become positive, and~\subref{fig:4d}
the heat capacity, (\ref{heat-cap1a33}), indicates that we obtained a negative heat capacity when $r_d<r_h$ and
the BH was unstable and when $r_d>r_h$ we have a positive heat capacity and the BH was stable.
}
\label{Fig:6}
\end{figure*}

The metric potential of the temporal component of Eq.~(\ref{elm3}) is given by
\begin{eqnarray}
\label{hor33}
&&f(r)=1-\frac{2M}{r}+\frac{q^2}{r^2}-\frac{q^4\mu}{10 r^6}-\frac{36q^4\mu^2}{7r^8}-\frac{140q^6\mu^3}{11r^{12}}\,.
\end{eqnarray}
Equation~(\ref{hor33}) is shown in Figure~\ref{Fig:6}~\subref{fig:6a}, which also indicates that the BH could possess two horizons at the root of $f(r)=0$.

The total mass contained within the event horizon ($r_h$) can be calculated by setting $f(r_h) = 0$.
Afterward the mass-radius of the horizon can be obtained as follows:
\begin{eqnarray}
\label{hor-mass-rad1a33}
&& {M_h}=\frac{770r_h{}^{10} \left( r_h{}^2+q^2 \right) - \left( 77r_h{}^6+3960r_h{}^4\mu+9800q^2\mu^2 \right)q^4\mu}{1540r_h{}^{11}}\, .
\end{eqnarray}
To present the aforementioned features differently, Figure~\ref{Fig:6}~\subref{fig:6b} depicts
value $M_h$, which corresponds to the $r_h$ horizon.

Using Eq.~(\ref{temp11}), the Hawking temperature can be calculated as:
\begin{eqnarray}
\label{T33}
&&T_h=\frac{2r_h{}^{10}(r_h{}^2-q^2)+q^4\mu(r_h{}^{4}[r_h{}^2+72\mu]+280\mu^2q^2)}{8\pi\,r_h{}^{13}}\,,
\end{eqnarray}
The behavior of the Hawking temperature given by Eq.~(\ref{T33}) is shown in Figure~\ref{Fig:6}~\subref{fig:6c} which indicates that
$T_h$ has a vanishing value at $r_h = r_d$ for different values of charge parameter $q$.

Substituting (\ref{hor-mass-rad1a33}) and (\ref{temp11}) into (\ref{heat-capacity11}), we obtain the heat capacity as follows:
\begin{equation}
\label{heat-cap1a33}
{H_h}
=\frac{2\pi r_h{}^2 \left[2r_h{}^{10} \left( r_h{}^2-q^2 \right) + q^4\mu \left[ r_h{}^{4} \left( r_h{}^2+72\mu \right) + 280\mu^2q^2\right] \right]}
{2r_h{}^{10} \left( 3q^2-r_h{}^2 \right) - q^4\mu \left[ r_h{}^{4} \left( 7r_h{}^2+648\mu \right) + 3640\mu^2q^2 \right]} \, .
\end{equation}
Equation~(\ref{heat-cap1a33}) shows that $H_h$ does not locally diverge and that the BH exhibits a phase transition of the second-order.
The heat capacity is depicted in Figure~\ref{Fig:4}~\subref{fig:4d} which also shows that $H_h<0$ when $\mu=-1$ for different values of charge parameter $q$.
The heat capacity is negative primarily because of the derivative of the Hawking temperature consistent with the nature of the MEH and Reissner Nordstr\"om BHs
which can be discovered at $\mu=0$.


\subsection{First law of thermodynamics of BH solutions (\ref{elm1}), (\ref{elm2}) and (\ref{elm3})}\label{fir}

An important step for any BH solution is to check its validity with the first law of thermodynamics.
Therefore, for the charged BH the Smarr formula and the differential form for the first law of
thermodynamics, in the frame of mimetic gravitational theory
can be expressed as follows \cite{Zheng_2018,Okcu:2017qgo}
\begin{equation}
\label{1st}
M(S,Q,\mu)=2(T\,S-\mathcal{A}\mu)+q(r)Q\,, \qquad \qquad dE=TdS+q(r) dQ+\mathcal{A}d\mu\, ,
\end{equation}
where $S$ is the Hawking entropy, $T$ is the Hawking
temperature, $q(r)$ is the electric potential and $\mathcal{A}$ is the conjugate of the Euler-Heisenberg parameter $\mu$.
Using Eqs.~(\ref{sol1}), (\ref{hor-mass-rad1a11}), (\ref{ent1}) and (\ref{T11}) in (\ref{1st}) we obtain
\begin{align}
\label{cnst}
\mathcal{A}=&\frac{q}{180r^5\mu^3}\left[27q^3\mu^2-90r^4q\mu+103^{2/3}\sqrt{\mu}
\mathlarger{\mathlarger{\mathlarger{\mathlarger{\int}}}} 
\frac{\sqrt[3]{9r^2\left(\sqrt{3\left( 8r^4+27q^2\mu \right)}-9q\sqrt{\mu}\right)^2}-6r^2}
{\sqrt[3]{r^4\left(\sqrt{3 \left( 8r^4+27q^2\mu \right)}-9q\sqrt{\mu} \right)^2}}dr\right]\nonumber\\
&\approx\frac{q^4}{10r^5}+\frac{\mu q^6}{24r^9}-\frac{3\mu^2 q^8}{52r^{13}}+ \mathcal{O}\left(\frac{1}{r^{17}}\right)+\cdots\,.
\end{align}
Using Eq.~(\ref{cnst}) in Eq.~(\ref{1st}), we can prove the first law of flat spacetime
(\ref{elm1}) and that first law (\ref{1st}) is verified for BH solutions (\ref{elm2}) and (\ref{elm3}).

\subsection{Stability of BHs (\ref{elm1}), (\ref{elm2}) and (\ref{elm3}) }\label{S5e}

The geodesic equations are given by
\begin{equation}
\label{ge3}
\frac{d^2 x^\alpha}{d\lambda^2}
+ \left\{ \begin{array}{c} \alpha \\ \beta \rho \end{array} \right\}
\frac{d x^\beta}{d\lambda} \frac{d x^\rho}{d\lambda}=0\, ,
\end{equation}
where $\lambda$ represents the affine connection parameter. The
geodesic deviation equations have the form \cite{1992ier..book.....D,Nashed:2003ee}
\begin{equation}
\label{ged333}
\frac{d^2 \epsilon^\sigma}{d\lambda^2}
+ 2\left\{ \begin{array}{c} \sigma \\ \mu \nu \end{array} \right\}
\frac{d x^\mu}{d\lambda} \frac{d \epsilon^\nu}{d\lambda}
+ \left\{ \begin{array}{c} \sigma \\ \mu \nu \end{array} \right\}_{,\, \rho}
\frac{d x^\mu}{d\lambda} \frac{d x^\nu}{d\lambda}\epsilon^\rho=0\, ,
\end{equation}
where $\epsilon^\rho$ is the four-vector deviation. Introducing (\ref{ge3}) and (\ref{ged333})
into (\ref{met}), we obtain
\begin{equation}
\label{ges}
\frac{d^2 t}{d\lambda^2}=0\, , \qquad \frac{1}{2} f'(r) \left(
\frac{d t}{d\lambda}\right)^2 - r\left( \frac{d \phi}{d\lambda}\right)^2=0\, , \qquad
\frac{d^2 \theta}{d\lambda^2}=0 \, ,\qquad \frac{d^2 \phi}{d\lambda^2}=0\, ,
\end{equation}
and for the geodesic deviation BH (\ref{met}) gives
\begin{eqnarray}
\label{ged11}
&& \frac{d^2 \epsilon^1}{d\lambda^2} +f_1(r)f'(r) \frac{dt}{d\lambda}
\frac{d \epsilon^0}{d\lambda} -2r f_1(r) \frac{d \phi}{d\lambda}\frac{d \epsilon^3}{d\lambda}
+\left[ \frac{1}{2} \left( f'(r)f'_1(r)+f_1(r) f''(r)
\right)\left( \frac{dt}{d\lambda} \right)^2-\left(f_1(r)+rf'_1(r)
\right) \left( \frac{d\phi}{d\lambda}\right)^2 \right]\epsilon^1=0\, ,
\nonumber\\
&& \frac{d^2 \epsilon^0}{d\lambda^2} + \frac{f'_1(r)}{f_1(r)} \frac{dt}{d\lambda}
\frac{d \zeta^1}{d\lambda}=0\, ,\qquad \frac{d^2 \epsilon^2}{d\lambda^2}
+\left( \frac{d\phi}{d\lambda}\right)^2 \epsilon^2=0\, , \qquad
\frac{d^2 \epsilon^3}{d\lambda^2} + \frac{2}{r} \frac{d\phi}{d\lambda}
\frac{d \epsilon^1}{d\tau}=0\, ,
\end{eqnarray}
where $f(r)$ and $f_1(r)$ are defined from Eq.~(\ref{met}) and
$'$ is derivative w.r.t. radial coordinate $r$. The use of a circular orbit gives
\begin{equation}
\label{so}
\theta= \frac{\pi}{2}\, , \qquad
\frac{d\theta}{d\lambda}=0\, , \qquad \frac{d r}{d\lambda}=0\, .
\end{equation}
Using Eq.~(\ref{so}) in Eq.~(\ref{ges}) yields
\begin{equation}
 \left( \frac{d\phi}{d\lambda}\right)^2={f'(r)
\over r[2f(r)-rf'(r)]}, \qquad \left( \frac{dt}{d\lambda}\right)^2= \frac{2}{2f(r)-rf'(r)}\, .
\end{equation}

\begin{figure}[ht]
\centering
\subfigure[~Stability of the BH (\ref{elm1})]{\label{fig:7a}
\includegraphics[scale=0.25]{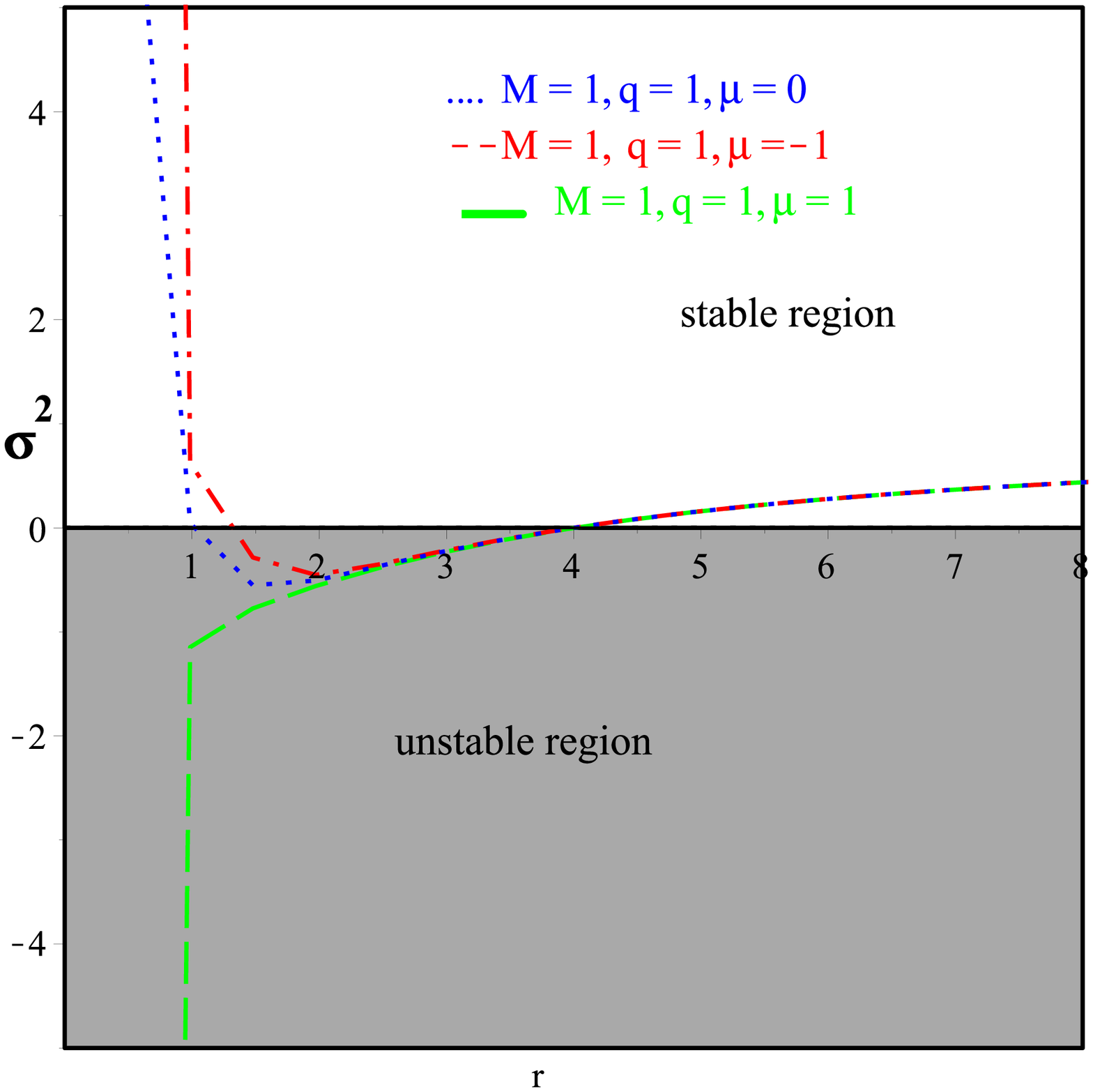}}\hspace{0.2cm}
\subfigure[~Stability of the BH (\ref{elm2})]{\label{fig:7b}
\includegraphics[scale=0.25]{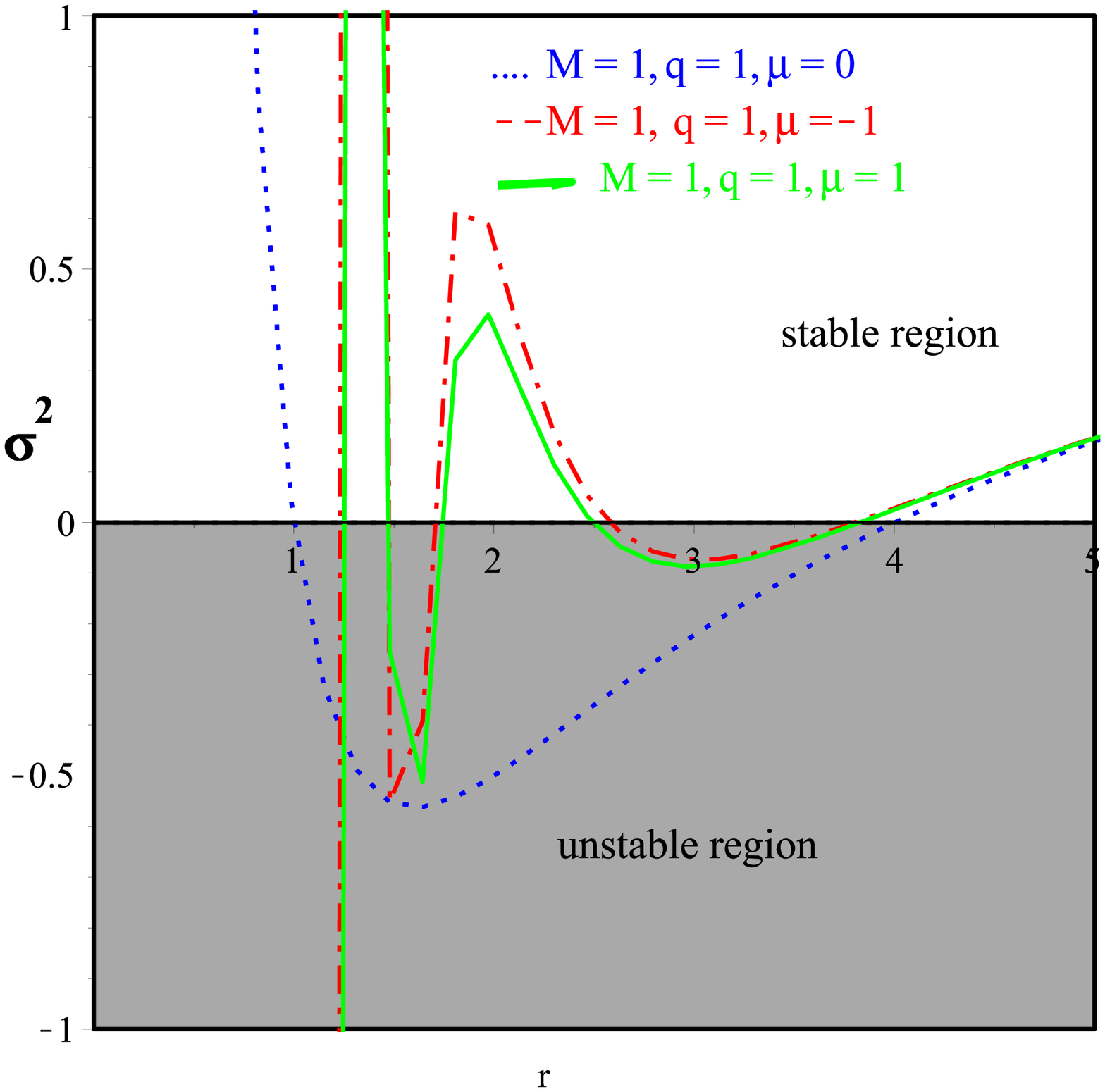}}
\subfigure[~Stability of the BH (\ref{elm3})]{\label{fig:7c}
\includegraphics[scale=0.25]{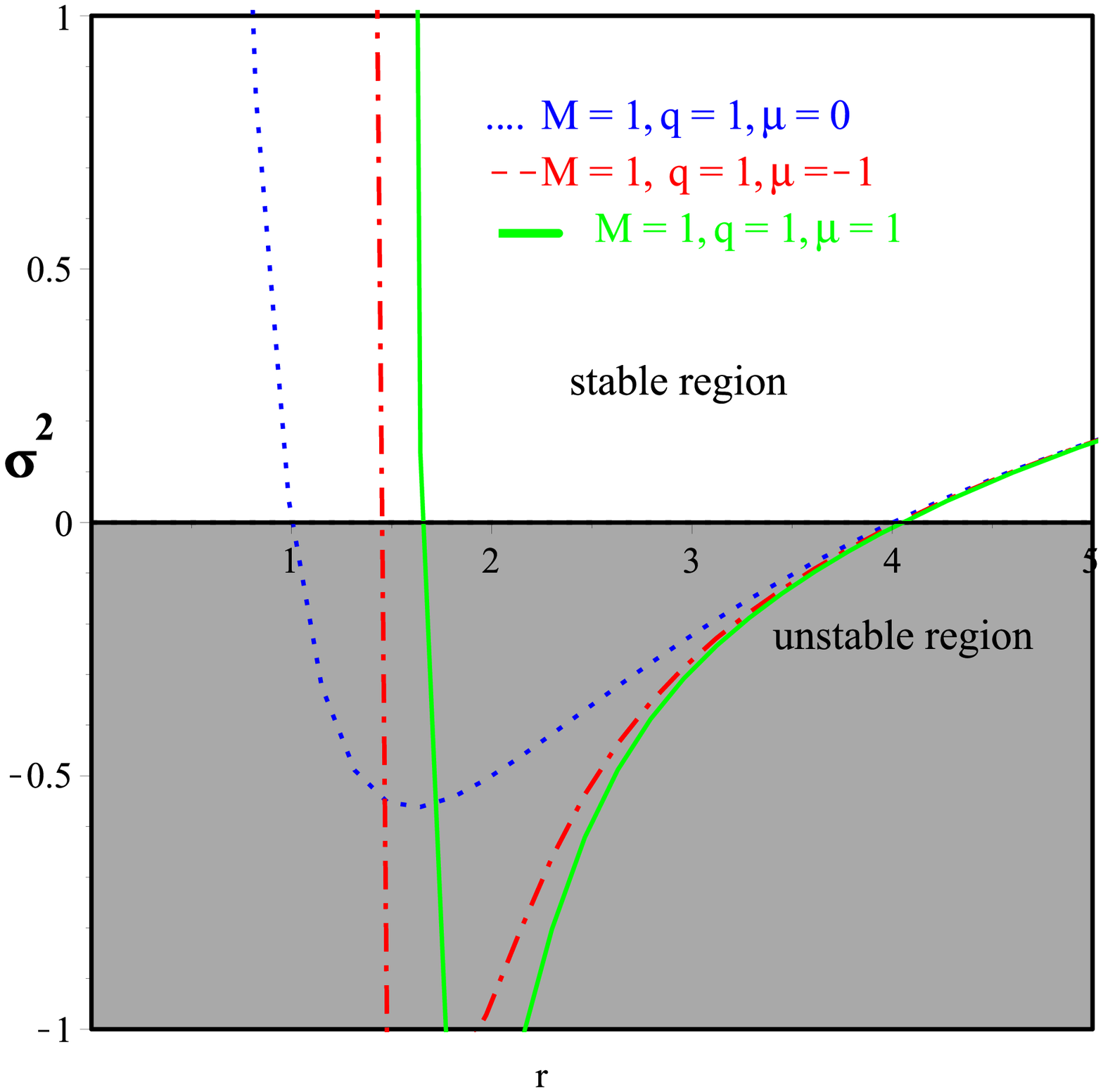}}
\caption{ {Plot of Eq.~(\ref{stab1}) against coordinate $r$ for BHs (\ref{elm1}), (\ref{elm2})
and (\ref{elm3}).}}
\label{Fig:7}
\end{figure}

The equations in (\ref{ged11}) can have the following form
\begin{eqnarray}
\label{ged22}
&& \frac{d^2 \epsilon^1}{d\phi^2}
+f(r)f'(r) \frac{dt}{d\phi} \frac{d \epsilon^0}{d\phi}
 -2r f_1(r) \frac{d \epsilon^3}{d\phi} +\left[ \frac{1}{2} \left[f'^2(r)+f(r) f''(r)
\right]\left( \frac{dt}{d\phi}\right)^2-\left[f(r)+rf'(r)
\right] \right]\zeta^1=0\, , \nonumber\\
&& \frac{d^2 \epsilon^2}{d\phi^2}+\epsilon^2=0\, , \qquad
\frac{d^2 \epsilon^0}{d\phi^2} + \frac{f'(r)}{f(r)}
\frac{dt}{d\phi} \frac{d \epsilon^1}{d\phi}=0\, , \qquad \frac{d^2 \epsilon^3}{d\phi^2}
+ \frac{2}{r} \frac{d \epsilon^1}{d\phi}=0\, .
\end{eqnarray}
 From the second equation of (\ref{ged22}) we can show that we have a simple harmonic
motion, i.e., the stability condition of plane $\theta=\pi/2$ providing that the rest of the equations in (\ref{ged22}) have solutions as follows
\begin{equation}
\label{ged33}
\epsilon^0 = \zeta_1 \e^{i \sigma \varphi}\, , \qquad
\epsilon^1= \zeta_2\e^{i \sigma \varphi}\, , \qquad \mbox{and} \qquad
\epsilon^3 = \zeta_3 \e^{i \sigma \varphi}\, ,
\end{equation}
where $\zeta_1$, $\zeta_2$ and $\zeta_3$ are constants and $\varphi$ is an unknown variable.
Using Eq.~(\ref{ged33}) in (\ref{ged22}), the stability condition for spacetime (\ref{met}) is:
\begin{equation}
\label{con111}
\frac{3f f_1 f'-\sigma^2f f'-2rf_1f'^{2}+rf_1 f f'' }{f f_1'}>0\, .
\end{equation}
Equation~(\ref{con111}) has the following solution
\begin{equation}
\label{stab1}
\sigma^2= \frac{3f f_1 f''-2rf_1f'^{2}+rf f_1 f'' }{f^2 f'_1{}^2}>0\, .
\end{equation}
Equation~(\ref{stab1}) is depicted in Figure~\ref{Fig:7} for the three cases of metric (\ref{met}) given by
Eqs.~(\ref{elm1}), (\ref{elm2}), and (\ref{elm3}) using particular values of the model.
\section{Multi-horizon solutions}\label{S6}
The simplest BH solution is described using the
Schwarzschild metric, where metric coefficient $g_{00}$ is
\begin{equation}
\label{hor111}
f= 1-\frac{2M}{r}=f_3(r)(r-r_1)\,,
\end{equation}
where $f_3(r)=\frac{1}{r}$ and $r_1=2M$. Equation~(\ref{hor111}) has only one horizon at $r=r_1=2M$ which is the event horizon.
Equation~(\ref{hor111}) can be generated from Eq.~(\ref{elm1}) when $\mu=q=0$. When $\mu=0$ and $q\neq 0$ we obtain the Reissner-Nordstr\"om in the form
\begin{equation}
\label{hor222}
f= 1-\frac{2M}{r}+\frac{q^2}{r^2}=f_4(r)(r-r_1)(r-r_2)\,,
\end{equation}
where $f_4(r)=\frac{1}{r^2}$, $r_1=M+\sqrt{M^2-q^2}$, and $r_2=M-\sqrt{M^2-q^2}$.
Equation~(\ref{hor222}) has two horizons at $r=r_1=M+\sqrt{M^2-q^2}$ and $r=r_2=M-\sqrt{M^2-q^2}$ which are the Cauchy and event horizons, respectively.
Equation~(\ref{hor111}) can be generated from Eq.~(\ref{hor222}) when $q=0$.
When $\mu\neq0$ and $q\neq 0$, we find the nonlinear electromagnetism in which there are six roots from which we can generate two real roots,
as discussed in Subsections~\ref{S5a}, \ref{S5b}, and \ref{S5c}, when nonlinear parameter $\mu$ has a negative value.
In this section, we can generate three real roots of Eq.~(\ref{elm1}) as follows:
\begin{equation}
\label{hor333}
f=1-\frac{2M}{r}+\frac{q^2}{r^2}-\frac{q^4\mu}{10 r^6}=f_5(r)(r-r_1)(r-r_2)(r-r_3)\,,
\end{equation}
where $r_1$ and $r_2$ are the Cauchy and event horizons, respectively, and $r_3$ is the radius horizon reproduced form the parameter $\mu$.
For Eq.~(\ref{hor333}) it is difficult to derive the explicit form of the three horizons.
Therefore, we will solve Eq.~(\ref{hor333}) numerically and graphically when $\mu>0$.
We plot Eq.~(\ref{hor333}) for specific values of the mass, charge, and parameter $\mu$ associated with the nonlinear electromagnetic field.

\begin{figure}[ht]
\centering
\subfigure[~Multi-horizon of the BH (\ref{elm1})]{\label{fig:8a}
\includegraphics[scale=0.35]{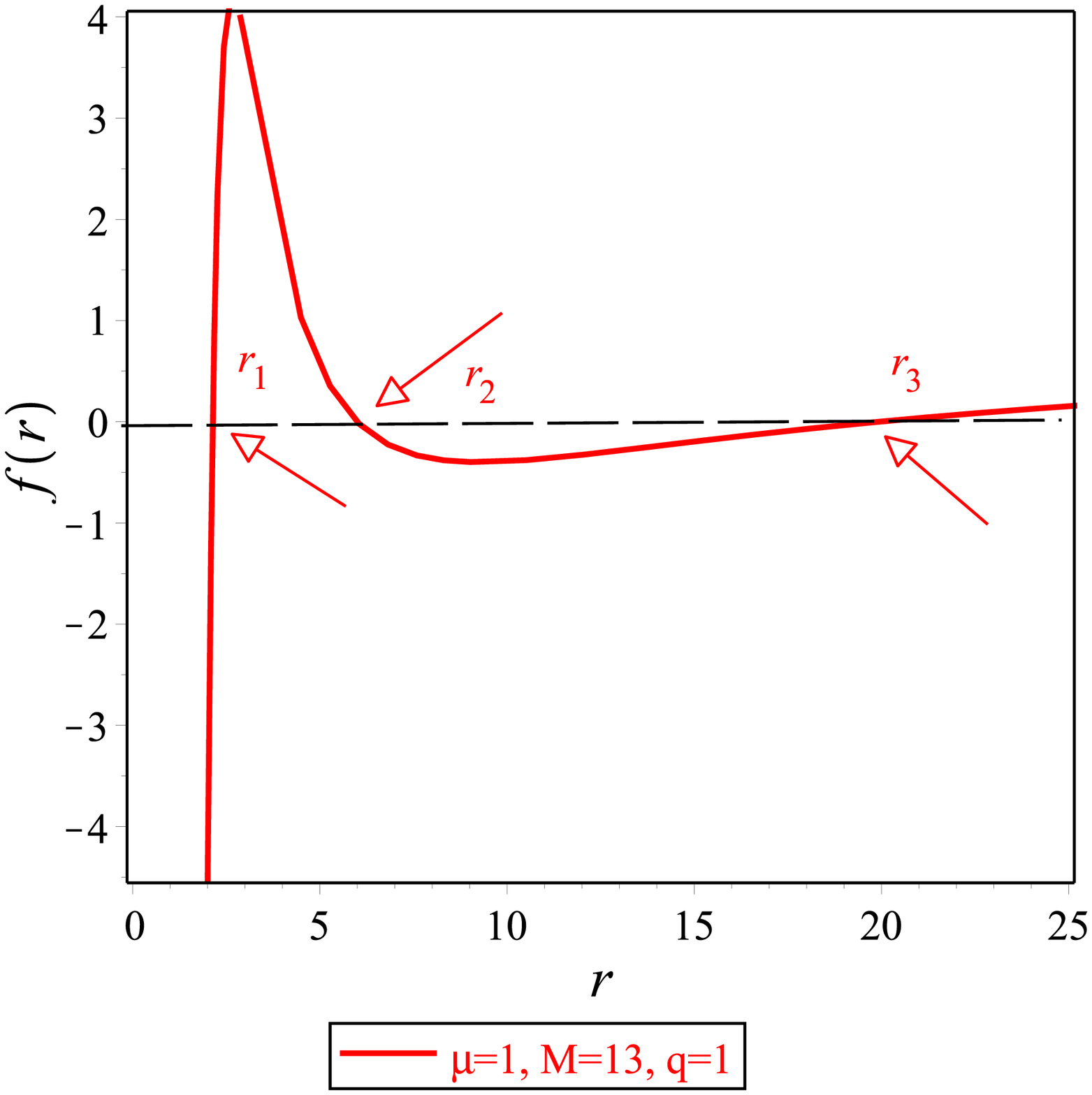}}\hspace{0.2cm}
\subfigure[~Multi-horizon of the BH (\ref{elm3})]{\label{fig:8b}
\includegraphics[scale=0.35]{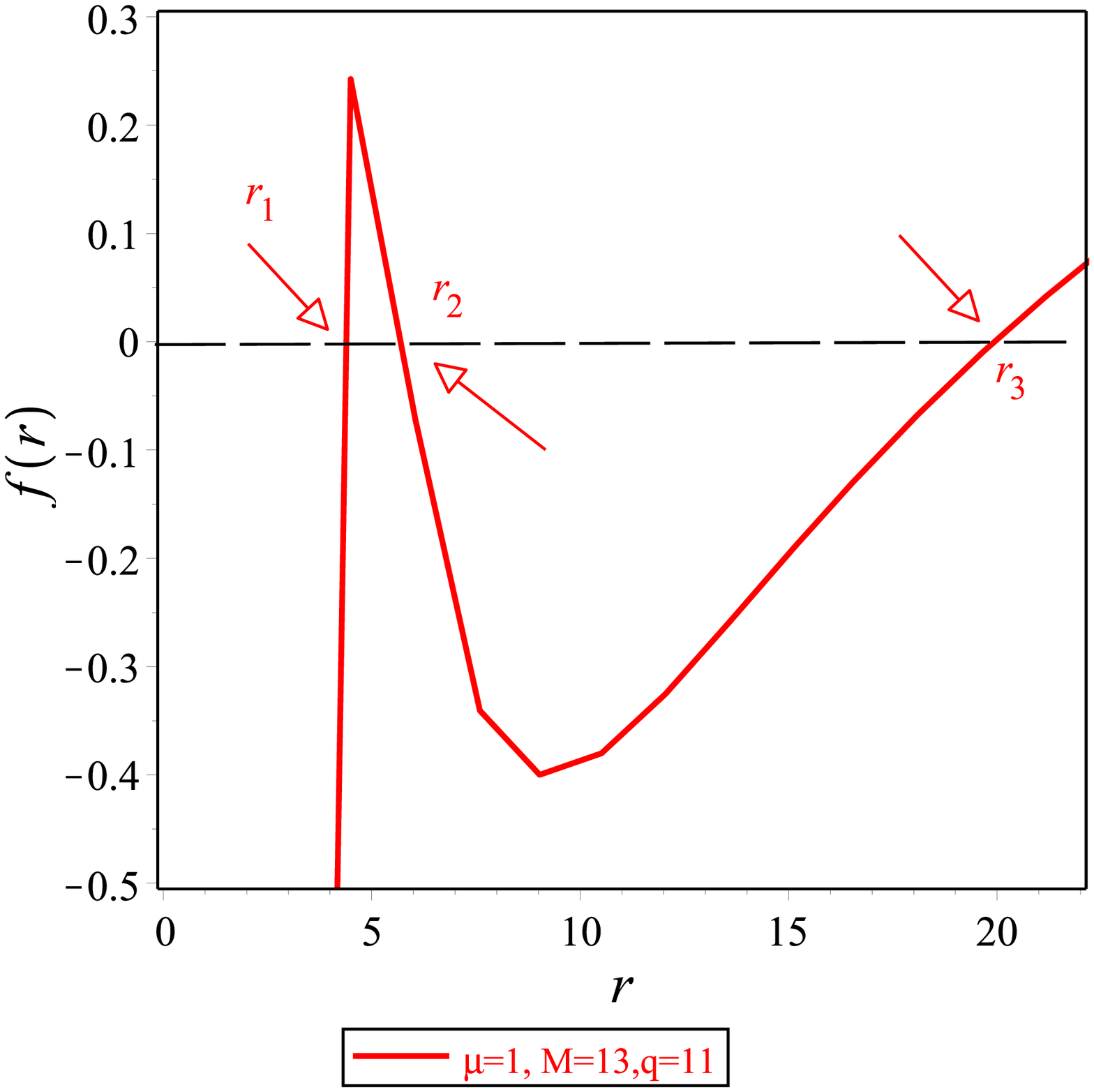}}
\caption{ {Multi-horizon plot of Eq.~(\ref{hor333}) against coordinate $r$ for BHs (\ref{elm1}), (\ref{elm3}).}}
\label{Fig:8}
\end{figure}
It is well-known that the curvature scalar of a BH that has one or two horizons is vanishing, the Schwarzschild and the Reissner-Nordstr\"om
but that the curvature scalar is singular in the center for a BH that has more than two horizons.
The Kretschmann scalar is singular in $r =0$ for any BH that has many horizons.
Equation~(\ref{inv}) clearly explains this.

\begin{figure*}
\centering
\subfigure[~The horizon Hawking temperature of the BH (\ref{elm1})]{\label{fig:9a}\includegraphics[scale=0.20]{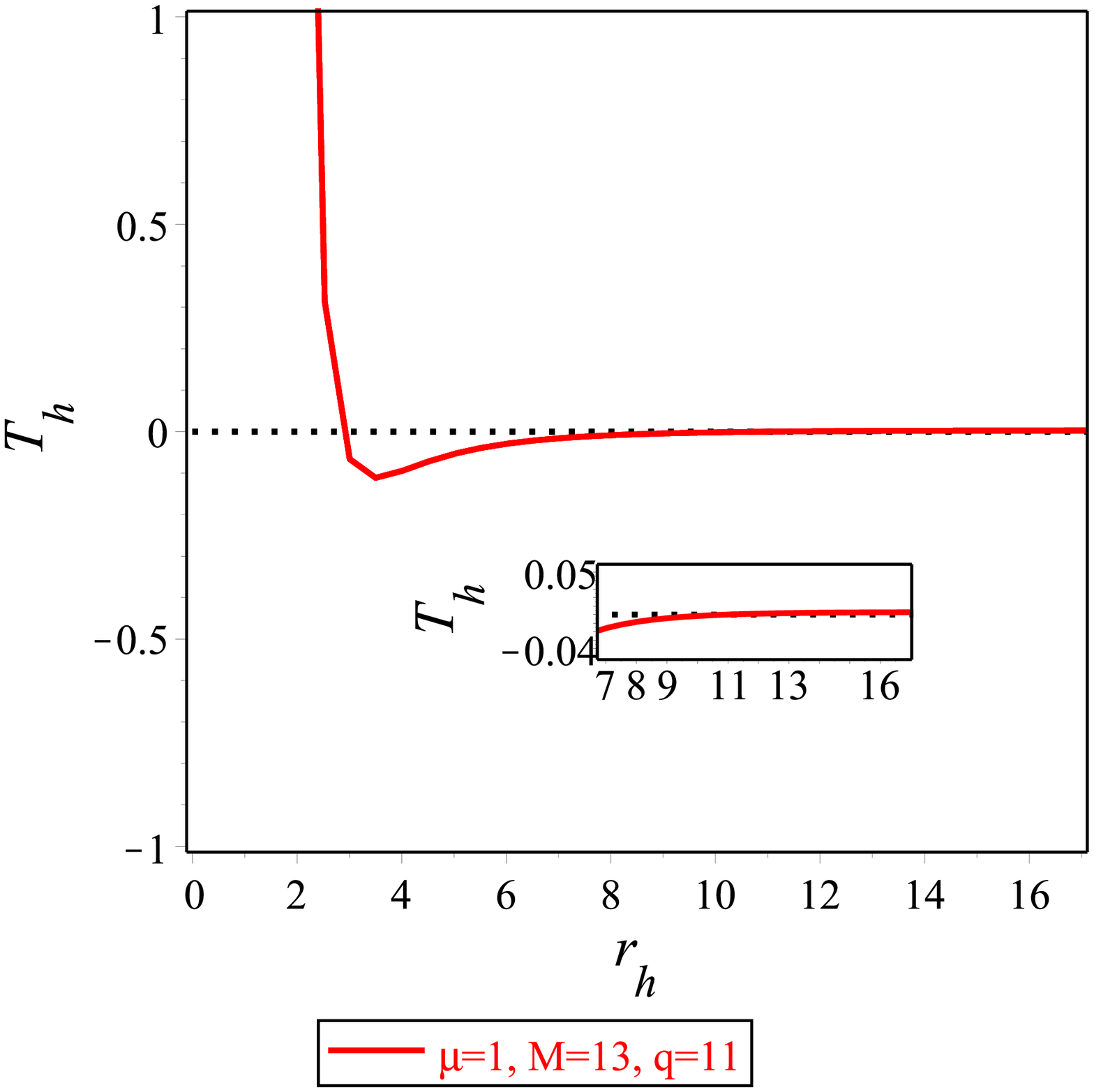}}\hspace{0.5cm}
\subfigure[~The horizon heat capacity of the BH (\ref{elm1})]{\label{fig:9b}\includegraphics[scale=0.20]{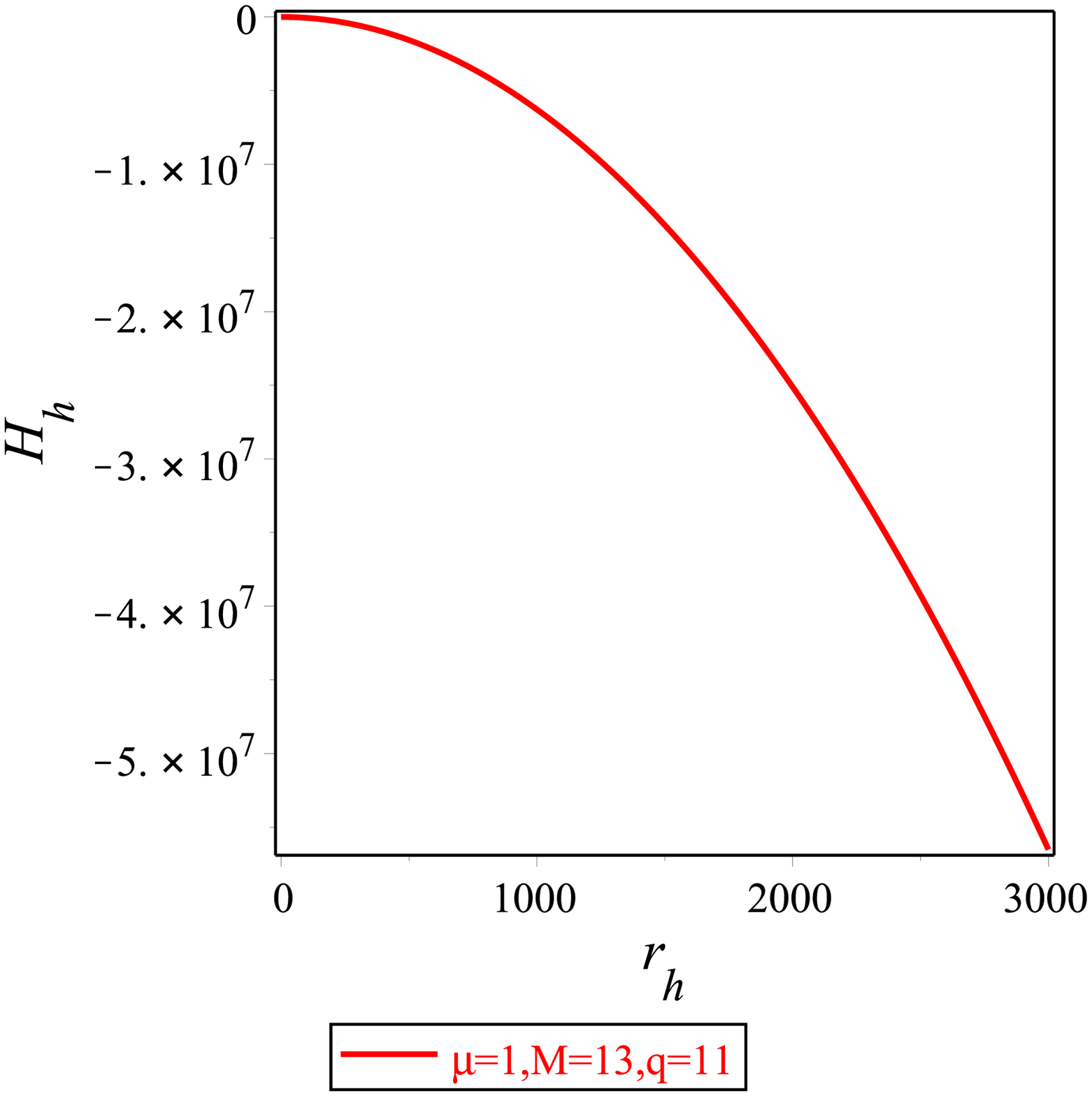}}\hspace{0.5cm}
\subfigure[~The horizon Hawking temperature of the BH (\ref{elm2})]{\label{fig:9c}\includegraphics[scale=0.20]{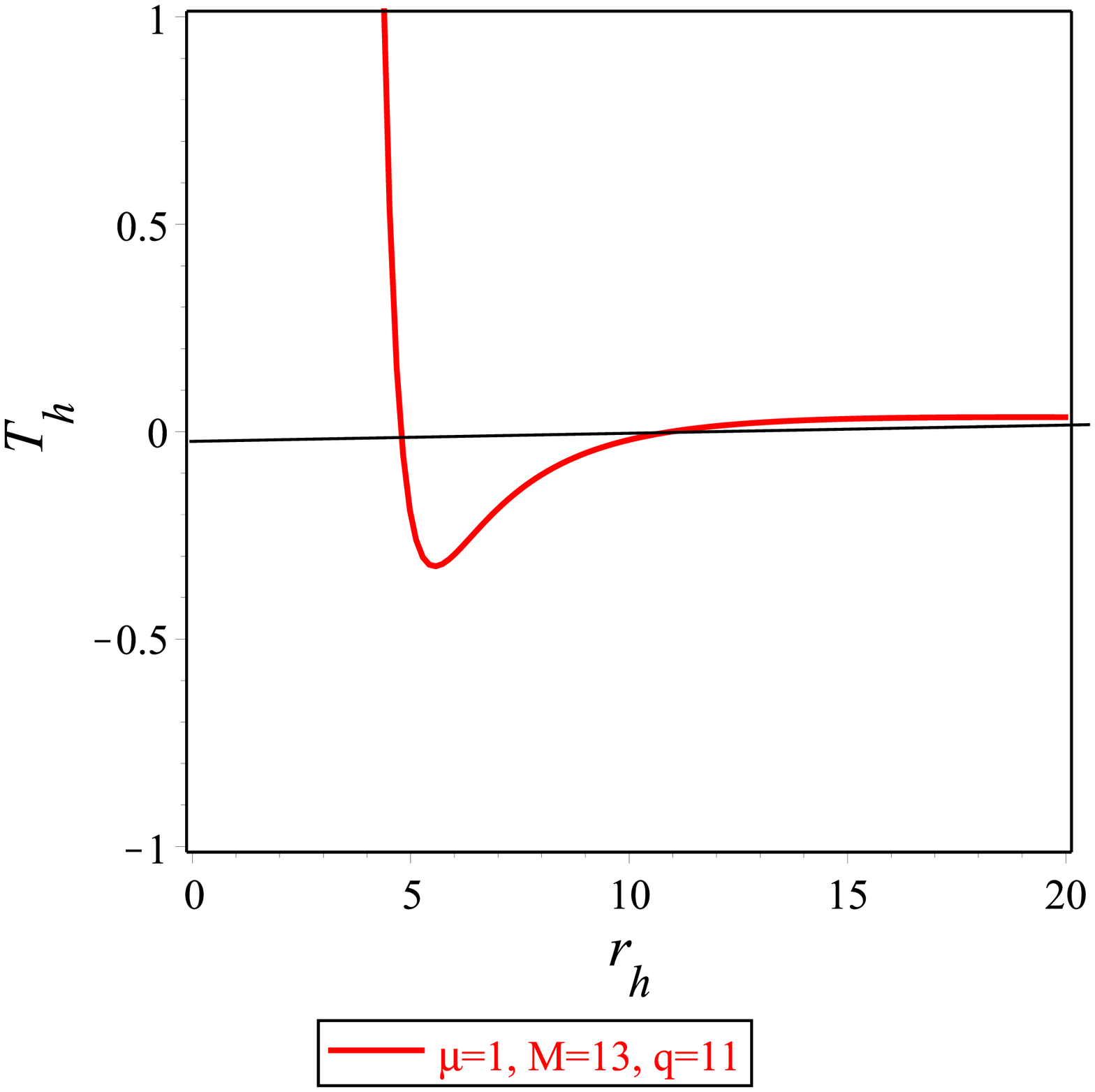}}\hspace{0.5cm}
\subfigure[~The horizon heat capacity of the BH (\ref{elm2})]{\label{fig:9d}\includegraphics[scale=0.20]{JMMCJMheatcapm}}\hspace{0.5cm}
\caption{Schematic plots of the thermodynamic quantities of the BH solutions (\ref{elm1}) and (\ref{elm2}) for the positive value of parameter $\mu$:~\subref{fig:6a}
typical behavior of the horizon Hawking temperature of the BH (\ref{elm1});~\subref{fig:6b} the horizon heat capacity of BH (\ref{elm1});~\subref{fig:6c}
typical behavior of the horizon Hawking temperature of BH (\ref{elm2}) and~\subref{fig:4d} the horizon heat capacity of BH (\ref{elm2}).
}
\label{Fig:9}
\end{figure*}

Now, let us study the thermodynamics of the aforementioned multi-horizons.
For BH (\ref{hor333}), we obtain the same quantities of thermodynamics presented in (Subsections (\ref{S5a}) and \ref{S5b})
but the behavior of these quantities differs because of the positive value of $\mu$.
All the plots of Figure~\ref{Fig:9} show that for a multi-horizon spacetime, we have unstable BHs.
\newpage
\section{Discussion}\label{S66}

Nowadays, most scientists believe the existence of the dark matter (DM)
because of the data stemming from observations of astrophysics and cosmology.
To explain this, scientists have proposed two directions to investigate dark matter,
to modify the field equations of Einstein's
GR or to modify the standard model by setting up new particle species.
Several studies have shown that these two directions are not different \cite{Sebastiani:2016ras}.
It is well-known that every amended gravity has new degrees of freedom in addition to the usual massless graviton of Einstein's GR.
As discussed in Section~\ref{S1} that the philosophy of mimetic gravity is to mimic DM, and it is a good candidate to explain the presence of cold dark matter.
Therefore, it is important to test the mimetic theory in the astrophysics domain, by explaining
possible novel BH solutions considering the EH term.

To carry out such a study, we delivered the equation of motion of the MEH gravitational theory and applied it to a four-dimensional spherically symmetric
metric with two unknown functions, $f(r)$ and $f_1(r)$, of radial coordinate $r$. We classified the metric into three cases:
Case I: $f(r)=f_1(r)$, Case II: $f(r)\neq f_1(r)$, and Case III: $f_1(r)=f(r)f_2(r)$.
Moreover, we used a vector potential with one unknown function related to the electric charge.
{ In this frame, we obtained charged BH solutions that included the mass, electric charge, and the EH parameter of the BH
and also a real positive value of the mimetic field.
In \cite{Myrzakulov:2015kda} and because of the form of the relevant equations, it had been recently realized that in a static spherically symmetric
spacetime, the mimetic field is assumed to be imaginary values, hence invalidating a direct connection with the degree
of freedom associated to dark matter.}
We also derived the mimetic field of each solution, which generally depended on the form of $f_1(r)$ and had a nontrivial value.
We showed that the asymptotic behavior of the BH solutions behaved as a flat spacetime.

We evaluated the invariants e.g., the Kretschmann $K=R_{\mu \nu \alpha \beta}R^{\mu \nu \alpha \beta}$,
the Ricci tensor squared $R_{\alpha \beta}R^{\alpha \beta}$,
and the Ricci scalar $R$, to investigate possible singularities of the derived BH solutions and founded that all BH solutions had true singularities at $r = 0$.
We demonstrated that MEH gravity produces singularities stronger than those in the Maxwell electrodynamics case.
Moreover, SEC and NEC were violated for all these BH solutions.
We evaluated the horizons and observed that when the EH took a negative value there were two horizons.
These BHs satisfied the first law of thermodynamics considering that the charge and EH parameter were variable.

We studied the possibility of multi-horizon BHs and showed that when the EH parameter, $\mu$, took a positive value,
we could generate three horizons. { Our BH solutions have three physical parameters, the mass $M$, charge $q$, and
the EH parameter $\mu$ which is responsible for the 3-horizon black hole. To our knowledge, this is the first time
to derive 3-horizon black holes with an analytic EH parameter.}
We researched the thermodynamics of these multi-horizon BH and found that their relevant temperature began
with a positive value became negative, and, finally became positive again.
Additionally, we demonstrated that the heat capacity of these multi-horizons is always negative, which means that such BH solutions are unstable.

{ Equation~(\ref{inv}) shows that the Ricci scalar of the BH solutions (\ref{sol1}), (\ref{sol2}) and (\ref{sol3}), is not constant,
which could mean that the solutions could not be equivalent to the solutions in the Einstein gravity
with a massless scalar field \cite{PhysRev.115.1325,Momeni:2015aea}.
We have shown that our BH solution has a naked singularity when the EH parameter has a positive value, as shown in Figure~\ref{Fig:4}~\subref{fig:4b},
however when this parameter has a negative value, the BH has a singularity, and we can have multi-horizons. }

{ An interesting point in cosmology is that the negative temperature and the positive heat capacity appear when the horizon radius $r_h$
is smaller than a critical radius $r_d$ as shown in Section~\ref{S4}.
In the early universe, many black holes might have generated by quantum fluctuations.
In general relativity, the smaller black holes have the higher Hawking temperatures, and therefore, the small black holes evaporate rapidly,
and they will disappear in the present universe.
In the solutions found in this study, however, even small black holes can have vanishing or negative temperature and positive heat capacity,
and therefore they do not evaporate and can remain even in the present universe.
Such primordial black holes might be a candidate for dark matter.
The production of the black holes in the early universe may be discussed in the future based on the mimetic Euler-Heisenberg theory. }

\begin{acknowledgments}
This work is supported by the JSPS Grant-in-Aid for Scientific Research (C)
No. 18K03615 (S.N.).
\end{acknowledgments}

%

\end{document}